\documentclass[a4paper,11pt]{article}
\pdfoutput=1
\usepackage{jheppub}
\usepackage{graphicx}
\usepackage{subfig}
\usepackage{amsmath}
\usepackage{amsfonts}
\usepackage{amssymb}
\usepackage{color}
\usepackage{dcolumn}

\usepackage[T1]{fontenc}
\title{\boldmath  Unbalanced St\"{u}ckelberg Holographic Superconductors with Backreaction}

\author[a]{Ahmad Jamali Hafshejani}
\author[b,c]{Seyed Ali Hosseini Mansoori}

\affiliation[a]{Physics Department, Yazd University, 89195-741,Yazd, Iran}

\affiliation[b]{Faculty of Physics, Shahrood University of Technology, P.O. Box 3619995161 Shahrood, Iran}
\affiliation[c]{
School of Astronomy, Institute for Research in Fundamental Sciences (IPM), P.O. Box 19395-5531, Tehran, Iran}

\emailAdd{ahmah.jamalii86@gmail.com, shosseini@shahroodut.ac.ir; shossein@ipm.ir}

\abstract{We numerically investigate some properties of unbalanced St\"{u}ckelberg holographic superconductors, by considering backreaction effects of fields on the background geometry. More precisely, we study the impacts of the chemical potential mismatch and St\"{u}ckelberg mechanism on the condensation and conductivity types (electrical, spin, mixed, thermo-electric, thermo-spin and thermal conductivity).
Our results show that the St\"{u}ckelberg's model parameters $C_{\alpha}$ and $\alpha$ not only have significant impacts on the phase transition, but also affect the conductivity pseudo-gap and the strength of conductivity fluctuations. Moreover, the effects of these parameters on a system will be gradually reduced as the imbalance grows.
We also find that the influence of $\alpha$ on the amplitude of conductivity fluctuations depends on the magnitude of the both $C_{\alpha}$ and $\delta\mu/\mu$ in the electric and thermal conductivity cases. This results in that increasing $\alpha$ can damp the conductivity fluctuations of an unbalanced system in contrast to balanced ones.}

\begin{document}
\maketitle
\flushbottom

\section {Introduction}

The \textit{gauge-gravity} duality \cite{Maldacena} based on the holographic principle, establishes a relationship between a gravitational theory in the bulk with $d+1$ dimensions and a quantum field theory on the boundary with $d$ dimensions. This duality can deal with lots of unsolved problems in strongly coupled field theories. One of the main achievements of this duality is the establishment of the holographic superconductors \cite{Hartnoll:2008vx, Hartnoll:2008kx, Herzog:2009xv,Gubser:2009qm}.

More precisely, the standard BCS theory \cite{Bardeen, Cooper}, which can describe the properties of low temperature superconductors, is not capable of fully explaining unconventional superconductors in strongly coupled regime. However, the \textit{gauge/gravity} duality may help us to handle strongly coupled systems and understand some features of the high temperature superconductors. This duality relies on the mechanism of spontaneously breaking of the global $U(1)$ symmetry in the dual field theory. This holographic model undergoes a phase transition from a black hole with no hair to a black hole with scalar hair at low temperatures \cite{Gubser:2008pf, Gubser:2008px}. There exist several studies on holographic superconductors to describe their different aspects \cite{Herzog:2010vz, Pan:2009xa, Cai:2010cv, Mansoori:2016zbp, Mahapatra:2013vta, Fan:2013tga, Sherkatghanad:2017edj, Jing:2010zp, Pan:2012jf, Hendi:2012zz}. 
One of the most interesting phenomena in superconductor research is the second order phase transitions in Abelian-Higgs models \cite{Gubser:2008px}.
Remarkably, the measurements of the ratio of pseduo-gap frequency to critical temperature ($\omega_{g}/T_c$) in standard holographic superconductors \cite{Hartnoll:2008vx} are in agreement with the experimental measurements of this ratio in the high temperature superconductors ($\omega_{g}/T_{c}\approx8 $) \cite{Gomes}. 

It is also interesting to take an effective field theory approach and consider the existence of the spontaneous symmetry breaking via the St\"{u}ckelberg mechanism \cite{Franco:2009yz, Franco:2009if, Aprile:2010yb}.
Such a model depends on a general function of the scalar field,  $\mathcal{F}(\psi)$.
One of the main features of this phenomenological model is provision of a large group of phase transitions which are the first order and second order phase transitions with non-mean field behavior.
In particular, the investigation of phase transitions in this model has achieved significant progress \cite{Pan:2010at, Pan:2011ns,Momeni,Ma:2011zze}.
Furthermore, in the conductivity case, additional resonances at non zero frequencies for some choices of function $\mathcal{F}$. One can interpret these poles as a sign of the existence of quasiparticles in the superconductor.
A similar behavior can be observed once the scalar field mass approaches the BF bound \cite{Horowitz:2008bn}.
In addition, St\"{u}ckelberg mechanisem enhances degrees of freedom (DOFs) of a given model by introducing a generic function $\mathcal{F}$ containing some parameters which can be fixed by experiments. Therefore, it is interesting to apply St\"{u}ckelberg mechanism in holographic superconductors to reach an efficient model with more DOF. The main purpose of this paper is to study the effects of this freedom in an unbalanced model \cite{Bigazzi:2011ak,Musso}.

An unbalanced model is based on an emerge of superconducting phase around a quantum critical point \cite{Sachdev:2011cs}. The mechanism of this model is that the superconductive phase happens where the two fermionic species contribute with unbalanced populations or unbalanced chemical potentials. This is a relevant subject both in condensed matter systems and QCD at finite density \cite{Casalbuoni:2003wh}. The unbalanced chemical potential can be produced by magnetic impurities in a system or by an existence of external magnetic field inducing Zeeman splitting of single-electron energy levels.

 In the holographic context, adding a non-trivial charged field on the gravity side leads to the breaking of a $U(1)_{A}$ ``charge'' symmetry which characterizes the onset of superconductivity \cite{Gubser:2008px, Hartnoll:2008vx, Hartnoll:2008kx}.
The chemical potential mismatch is also a potential for a $U(1)_{B}$ ``spin'' symmetry under which the scalar field is uncharged \cite{Iqbal:2010eh}.
These two gauge fields correspond to two conserved currents in the boundary theory which provides us with the strong-coupling generalization of the two-current model proposed by Mott \cite{Mott}. Furthermore, mixing effects of these two currents creates the spintronic features. One can, therefore, investigate the mixed spin-electric linear response by using the holographic method \cite{Bigazzi:2011ak, Musso}.
In Ref. \cite{Larkin, Fulde}, Larkin, Ovchinnikov, Fulde, and Ferrel showed that, except for the normal/superconductor phase transition, a system may also experience a new state called LOFF phase.
This inhomogeneous phase with spatially modulated condensate leads to spontaneously non-trivial spatial modulations.
Since St\"{u}ckelberg mechanism results in various phase transitions, its mixture with an unbalanced model can provide us with an appropriate theory to search for inhomogeneous superconducting phases.

In this paper, we study an unbalanced St\"{u}ckelberg holographic superconductor where the backreaction effects of matter on the geometry has been considered. In other words, we going to investigate the behaviors of holographic St\"{u}ckelberg superconductors in Ref. \cite{Franco:2009yz, Franco:2009if} in the presence of an imbalance.
Or equally, we look for how behaviors of unbalanced systems obtained in \cite{Bigazzi:2011ak, Musso} are affected by applying the St\"{u}ckelberg mechanism.
This mechanism is characterized by a generic function ${\cal F}(\psi)$ and goes to the Higgs mechanism by setting ${\cal F}(\psi)=\psi ^ 2$.
Therefore, in order to trace the effects of the St\"{u}ckelberg mechanism and imbalance on all types of conductivity, we need to construct the conductivity matrix describing the linear response of the system to variations of the external sources. In most cases, results show that the imbalance makes the influences of the St\"{u}ckelberg mechanism weaker. However, diagrams illustrate complicated behaviors in some situations.

The paper is organized as follows. In Section \ref{sec2}, we introduce the Lagrangian for our model. We also numerically calculate condensation and phase transition for different values of ${\delta\mu}/{\mu}$ (the ratio of the chemical potential mismatch to the chemical potential where indicates the amount of the imbalance) and ${\cal F}(\psi)$. In section \ref{sec3}, we briefly introduce the process of calculations for all types of the conductivity.
Then, we verify their response to changes in the form of ${\cal F}(\psi)$ function and the imbalance. Finally, conclusions are presented in Section \ref{sec4}.

%%%%%%%%%%%%%%%%%%%%%%%%%%%%%%%%%%%%%%%%
\section {The Model}\label{sec2}
%%%%%%%%%%%%%%%%%%%%%%%%%%%%%%%%%%%%%%%%

We consider an extension of the generalized St\"{u}ckelberg model introduced in \cite{Franco:2009yz} in which an extra $U(1)$ gauge field $B$ is added in the bulk. This gauge field is dual to spin current in the boundary theory. Note that the scalar field $\psi$ is uncharged under the additional gauge field $B$. Therefore, the bulk action for such an unbalanced St\"{u}ckelberg model in (3+1)-dimensions is defined as:
\begin{equation}\label{S}
 S = \frac{1}{2\kappa_4^2} \int dx^4 \sqrt{-g}
 \left({\cal R} + \frac{6}{L^2}
 + {\cal L}_{\rm{matter}} \right) \ ,
\end{equation}
where
\begin{equation}\label{L}
 {\cal L}_{\rm{matter}}=
  -\frac{1}{4} F^2
  -\frac{1}{4} Y^2
  -V(|\psi|)
  -(\partial \psi)^2
  -{\cal F}(\psi)(\partial p-q A)^2 
\end{equation}
in which $F=dA$ and $Y=dB$ are the two field strengths associated with the two gauge fields. The Maxwell equation makes the phase of $\psi$ constant, so we take it to be null in order to have real $\psi$. In addition, this theory is invariant under the local gauge symmetry $A\rightarrow A+\partial\Omega(x)$ and $p\rightarrow p+\Omega(x)$ \cite{Franco:2009yz}. Therefore, we can utilize the gauge freedom to fix $p=0$. We also set $L=1$ and $2\kappa_4^2=1$. Moreover, function ${\cal F}(\psi)$  can be written in a general form as:
 \begin{equation}\label{F}
 {\cal F}(\psi)=\psi^2+C_\alpha \psi^\alpha.
\end{equation}
It is obvious that our model reduces to the unbalanced model in Ref. \cite{Bigazzi:2011ak, Musso} when ${\cal F}(\psi)=\psi^2$. Note that we should take this function to be positive because of the positivity of the kinetic term.
The properties of the CFT at the boundary can change under the influence of function $\mathcal{F}(\psi)$ \cite{Franco:2009if}. In the effective field theory context, a change in the form of this function can correspond to a sort of ``non normalizable deformation'' or, equivalently, a change in the theory.

A plane-symmetric black hole, with considering backreaction effects, can be described by the metric ansatz: 
\begin{equation}\label{ansatz1}
 ds^2 = -g(r) e^{-\chi(r)} dt^2
 + r^2 (dx^2 + dy^2)
 + \frac{dr^2}{g(r)}\ .
\end{equation}
We also consider the following ansatz for the scalar and the vector fields:
\begin{equation}\label{ansatz2}
 \psi = \psi(r)\ , \ \ \ \
 A_a\, dx^a = \phi(r)\, dt\ , \ \ \ \
 B_a\, dx^a = v(r)\, dt\ .
\end{equation}
Furthermore, the temperature of such a black hole with the horizon at $r=r_h$ is defined as:
\begin{equation}\label{temp1}
T=\frac{g^\prime(r_h)e^{-\chi(r_h)/2}}{4\pi}\,.
\end{equation}
By varying the action with respect to the metric and the fields, we arrive at the following equations of motions,
\begin{equation}\label{EOM1}
 \psi'' + \psi' \left(\frac{g'}{g} + \frac{2}{r} - \frac{\chi'}{2}\right)
 - \frac{V'(\psi)}{2 g}
 + \frac{e^\chi q^2 \phi^ 2 \dot{{\cal F}}(\psi)}{2 g^2} = 0\,,
\end{equation}
\begin{equation}\label{EOM2}
 \phi'' + \phi'\left( \frac{2}{r} + \frac{\chi'}{2} \right)
 - \frac{2 q^2 {\cal F}(\psi)}{g} \phi = 0\,,
\end{equation}
\begin{equation}\label{EOM3}
 \frac{1}{2} \psi'^2
 + \frac{e^\chi (\phi'^2 + v'^2)}{4g}
 + \frac{g'}{gr}
 +\frac{1}{r^2}
 -\frac{3}{g}
 + \frac{V(\psi)}{2g}
 + \frac{e^\chi q^2 {\cal F}(\psi) \phi^2}{2g^2} = 0\,,
\end{equation}
\begin{equation}\label{EOM4}
 \chi' + r\psi'^2
 + r \frac{e^\chi q^2 \phi^2 {\cal F}(\psi)}{g^2} = 0\,,
\end{equation}
\begin{equation}\label{EOM5}
 v'' + v'\left( \frac{2}{r} + \frac{\chi'}{2} \right) = 0\,,
\end{equation}
where the prime denotes derivative with respect to $r$ and the dot denotes derivative with respect to $\psi$.
We also take the standard choice of mass as $m^2=-2$ \cite{Gauntlett:2009bh, Bobev:2011rv} and restrict the potential to $V(\psi)= m^2 \psi^2$ containing just the mass term. For our case, in which $m^2>-9/4$, the Breitenlohner-Freedman (BF) bound \cite{Breitenlohner} is respected.

In order to solve the set of equations \eqref{EOM1}-\eqref{EOM5}, one needs to impose suitable boundary conditions at the horizon and AdS boundary.
The asymptotic behavior of the scalar and gauge fields near the AdS boundary, $r\to \infty$, are: 
\begin{equation}\label{psi large r}
 \psi(r) = \frac{\psi_1}{r} + \frac{\psi_2}{r^2} + ...,
\end{equation}
\begin{equation}
 \phi(r) = \mu - \frac{\rho}{r} + ... \ , \ \ \ \
 v(r) = \delta\mu - \frac{\delta\rho}{r} + ...,
\end{equation}
where $\psi_{1}$ ($\psi_{2}$) can be regarded as the source of the dual condensation operator, ${\cal O}_{1}$ (${\cal O}_{2}$). Since we need the $U(1)$ symmetry to be broken spontaneously, we should turn one of the sources off. Therefore, we set $ \psi_1 = 0$ and $\langle {\cal O}_2 \rangle = \sqrt{2}\ \psi_2 $.
According to the \textit{gauge/gravity} duality, the leading terms of $\phi(r)$ ($v(r)$) are interpreted as chemical potential (chemical potential mismatch) and charge density (charge density mismatch) in the dual theory, respectively. Working in the grand-canonical ensemble, we fix the chemical potential (chemical potential mismatch) and alter the charge density (charge density mismatch).
At the AdS boundary, we also should set $\chi=0$ and impose the asymptotic behavior 
\begin{equation}
      g(r)=r^2-\frac{\epsilon}{2r}+...,
\end{equation}
where $\epsilon$ is the mass of black hole interpreted as the energy density of the dual field theory \cite{Hartnoll:2008kx}.

The other boundary conditions are those which are imposed at the horizon, $r=r_h$.
In this region, both $g(r)$ and the temporal component of the gauge fields vanish. Therefore, we have
\begin{equation}
 g(r_h)=\phi(r_h) = v(r_h)=0.
\end{equation}
By substituting Taylor expansion of fields at horizon in \eqref{temp1} and making use of the Einstein equation \eqref{EOM3}, the black hole temperature can be rewritten as
\begin{equation}\label{temp2}
 T = \frac{r_h}{16 \pi} \left[e^{-\frac{\chi_{h0}}{2}}
 \left(12 - 2 m^2 \psi^2_{h0}\right)
 - e^{\frac{\chi_{h0}}{2}} \left(\phi^2_{h1} + v^2_{h1}\right)\right]\ ,
\end{equation}
where subindexes $h0$ and $h1$ indicate the coefficients of the field's expansion at $r=r_h$.

Both the bulk and the boundary theory have the same time coordinate and, consequently, they have the same complex time continuation and temperature.
We numerically solve the equations of motion (\eqref{EOM1}-\eqref{EOM5}) by integrating from the horizon out to the infinity with respect to the mentioned boundary conditions.
We mostly consider the interval $0\leq\delta \mu/\mu\leq2$ with a fixed chemical potential, $\mu=1$. 
%%%%%%%%%%%%%%%%%%%%%%%%%%%%%%%%%%%%%%%%%%%%%%%%%%%%%%%%%%

\begin{figure}[t]
\centering
  \includegraphics{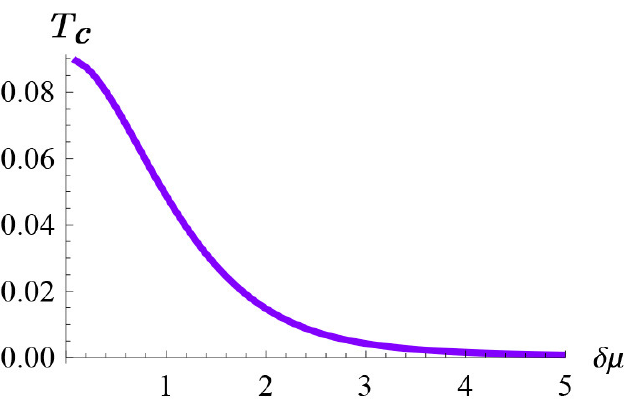}
    \caption{Diagram of critical temperature $T_c$ as a function of $\delta \mu$ by considering~\eqref{F} with $\alpha>2$.}
     \label{Tcdm}
\end{figure}

\begin{figure}[h]
\centering
  \includegraphics{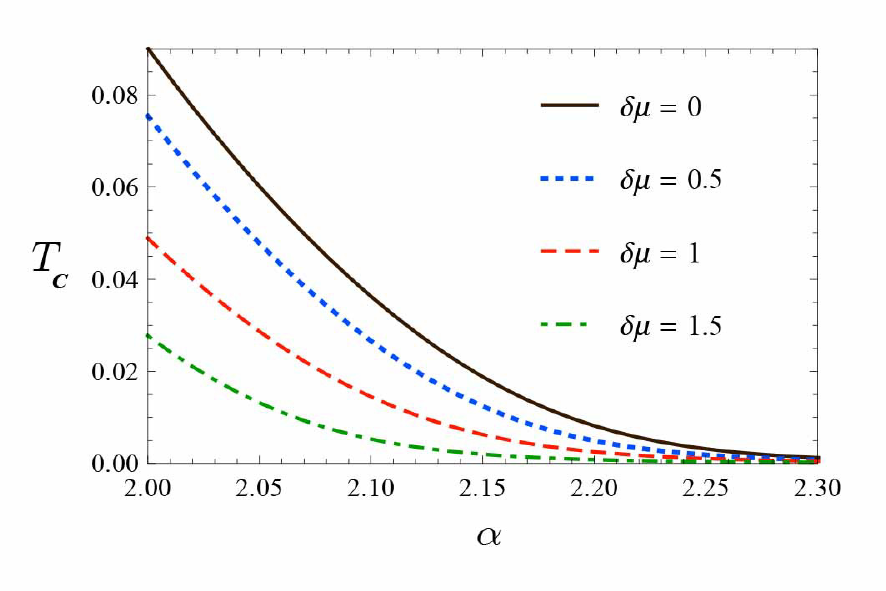}
  \caption{Diagram of critical temperature $T_c$ as a function $\alpha$ for the chosen function ${\cal F}(\psi)=\psi^\alpha$. From up to down we have $\delta\mu=0,0.5,1,1.5$.}
    \label{Tcalpha}
\end{figure}

\begin{table}[hbt]
\centering
  \begin{tabular}
   {|>{\centering\arraybackslash}m{1em}|>{\centering\arraybackslash}m{3em}|>{\centering\arraybackslash}m{3em}|>{\centering\arraybackslash}m{3em}|>{\centering\arraybackslash}m{3em}|}
   \hline
   $\alpha$ & 2 & 2.1 & 2.2 & 2.3 \\
   \hline
   $T_c$ & 0.0488 & 0.0145 & 0.0025 & 0.0005 \\
   \hline
  \end{tabular}
 \caption{Value of critical temperature $T_c$ as a function of $\alpha$ for given ${\cal F}(\psi)=\psi^\alpha$ and fixed $\delta\mu=1$.}
 \label{table:Tc}
\end{table}

\subsection{Condensation and phase transition}

In this subsection, we are looking for phase transition properties through study of the condensation of the scalar operator. 
Firstly, we plot the second order phase transition diagrams in the $(T_c,\delta \mu)$ plane for ${\cal F(\psi)}=\psi^2+C_\alpha \psi^\alpha$, with $\alpha >2$ and $\mu=1$. From Fig. (\ref{Tcdm}), we find that the critical temperature is not affected by the parameters in \eqref{F} since $\alpha >2$.
Of course, it could be predictable since at limit $\psi \to 0$ (near the normal phase), the dominant term in the function ${\cal F}(\psi)$ is $\psi^2$.
While, if we assume, for instant,  ${\cal F}(\psi)=\psi^\alpha$ from Ref. \cite{Franco:2009yz}, the critical temperature will be affected by $\alpha$ change. We check numerically this assertion by plotting $T_c$ with respect to $\alpha$ for the function $\cal F(\psi)=\psi^\alpha$ and various values of $\delta \mu$ parameter in Fig. (\ref{Tcalpha}). These curves explicitly show the $T_c$ dependence on $\alpha$ as well as $\delta \mu$. However, for our model in which function \eqref{F} with $\alpha >2$ is considered, the $T_c$ is only affected by $\delta \mu / \mu$.
We also represent some data in table (\ref{table:Tc}) which indicates the $\alpha$-dependence of $T_c$ for ${\cal F}(\psi)=\psi^\alpha$ and $\delta \mu / \mu=1$. In the following we consider a few forms of function $\cal F(\psi)$ and investigate phase transitions.

%%%%%%%%%%%%%%%%%%%%%%%%%%%%%%%%%%%%%%%%%%%%%%%%%%%%%%%%%%%%

\subsubsection{The case of ${\cal F}(\psi)=\psi^2+C_4 \psi^4$}

\begin{figure}[h]
    \centering    
        \subfloat[]{\includegraphics[width=0.5\columnwidth]{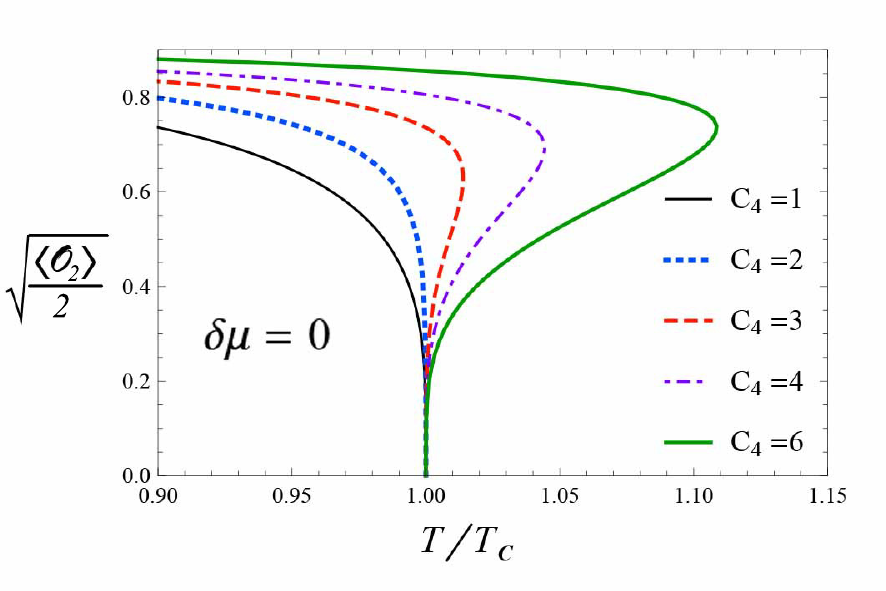}}
        \subfloat[]{\includegraphics[width=0.5\columnwidth]{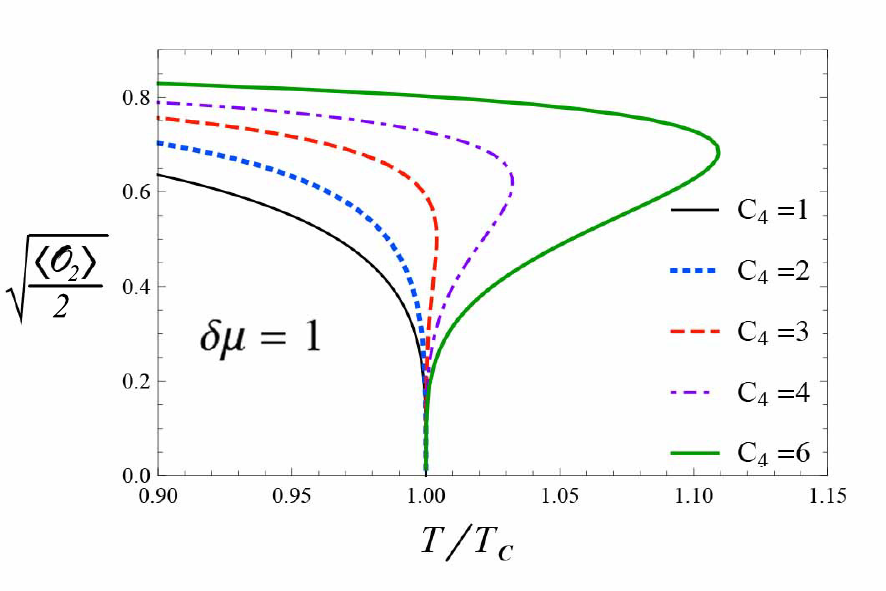}}
        \qquad
        \subfloat[]{\includegraphics[width=0.5\columnwidth]{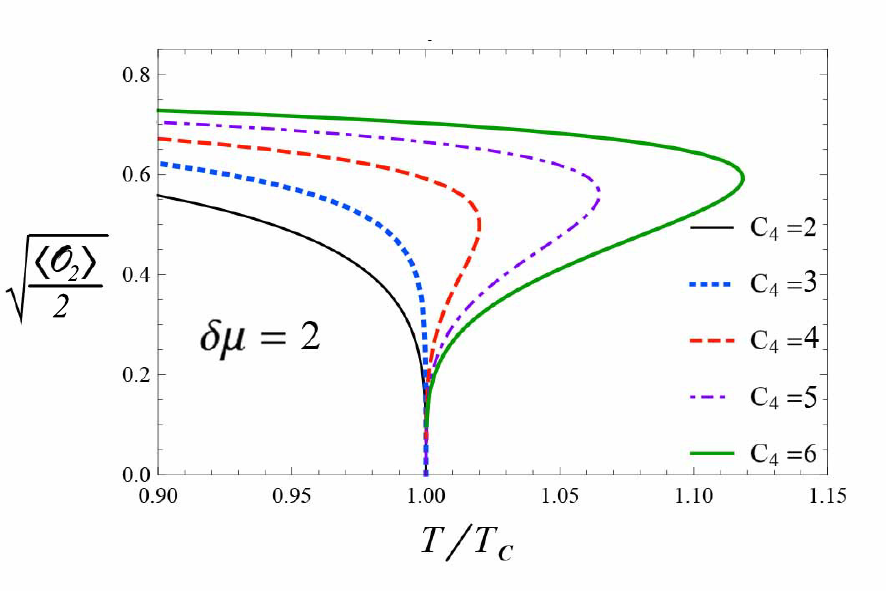}}
        \subfloat[]{\includegraphics[width=0.5\columnwidth]{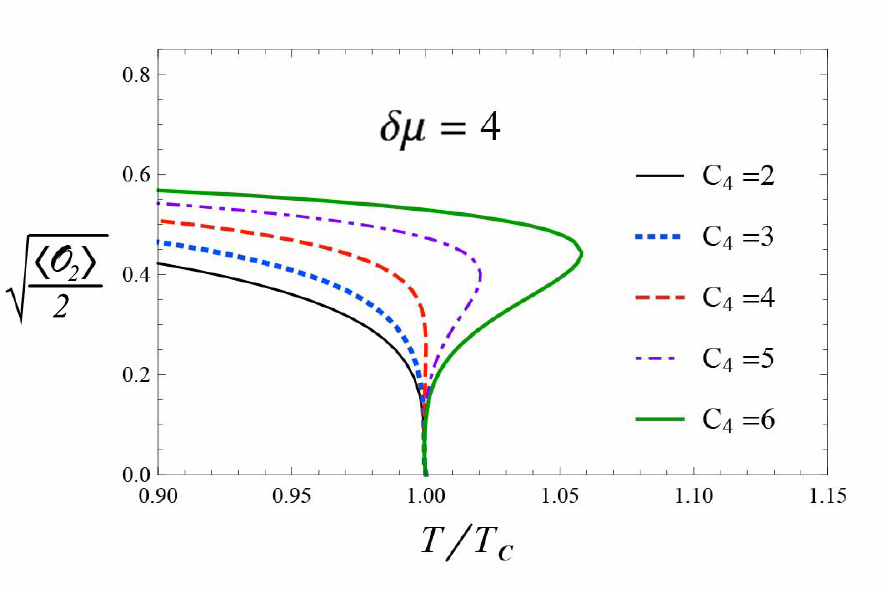}}
    \caption{Condensation versus temperature normalized by $T_c$ for chosen function ${\cal F}(\psi)=\psi^2+C_4 \psi^4$. }
        \label{CS2C4}
\end{figure}

\begin{figure}[h]
    \centering    
        \subfloat[]{\includegraphics[width=0.5\columnwidth]{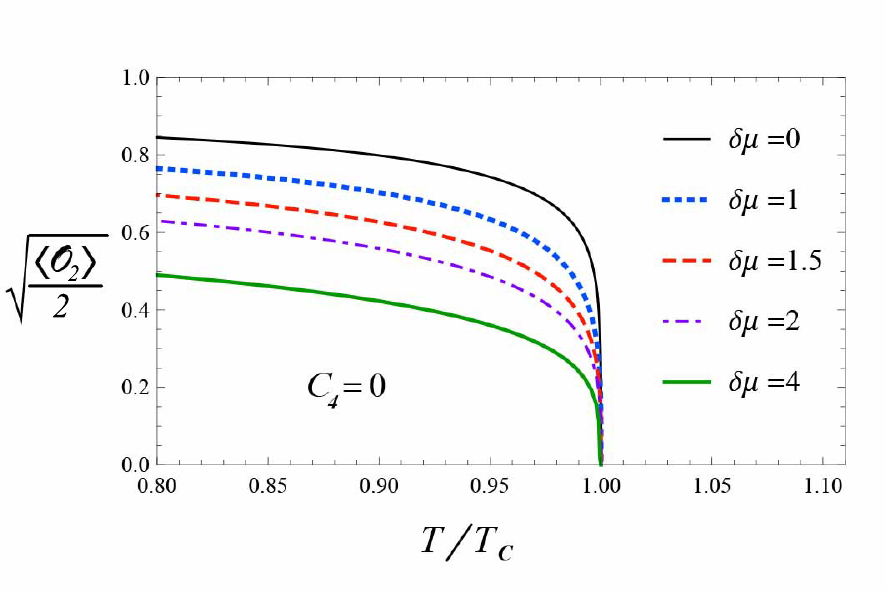}}
        \subfloat[]{\includegraphics[width=0.5\columnwidth]{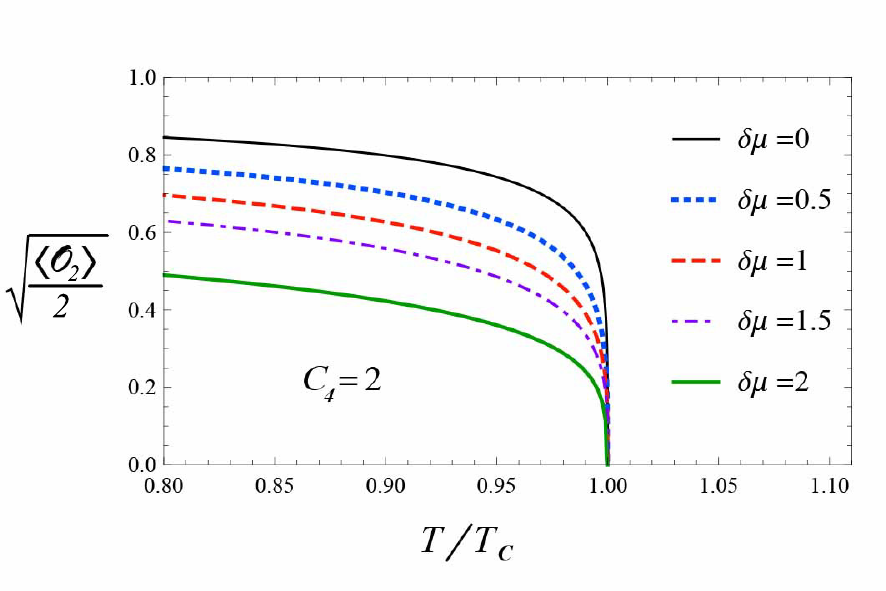}}        
       \qquad
       \subfloat[]{\includegraphics[width=0.5\columnwidth]{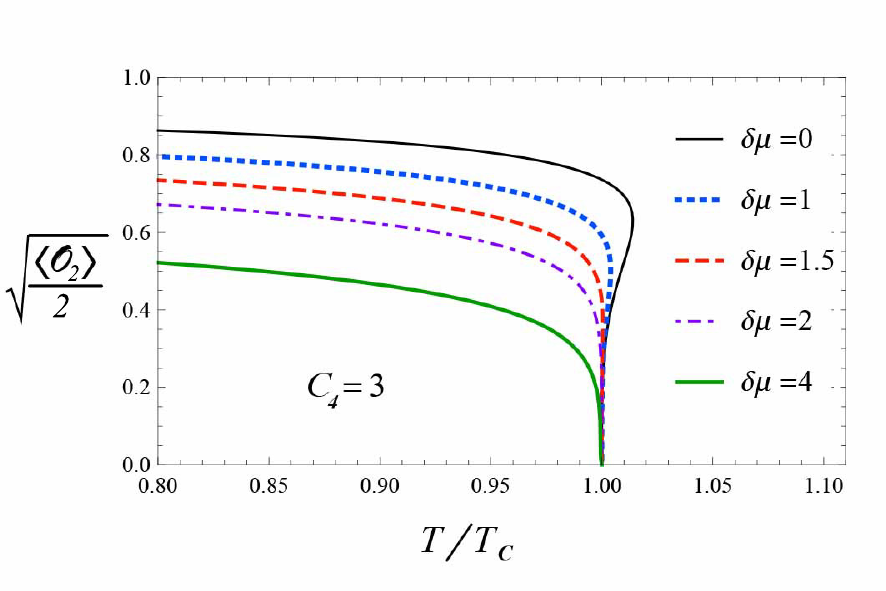}}
        \subfloat[]{\includegraphics[width=0.5\columnwidth]{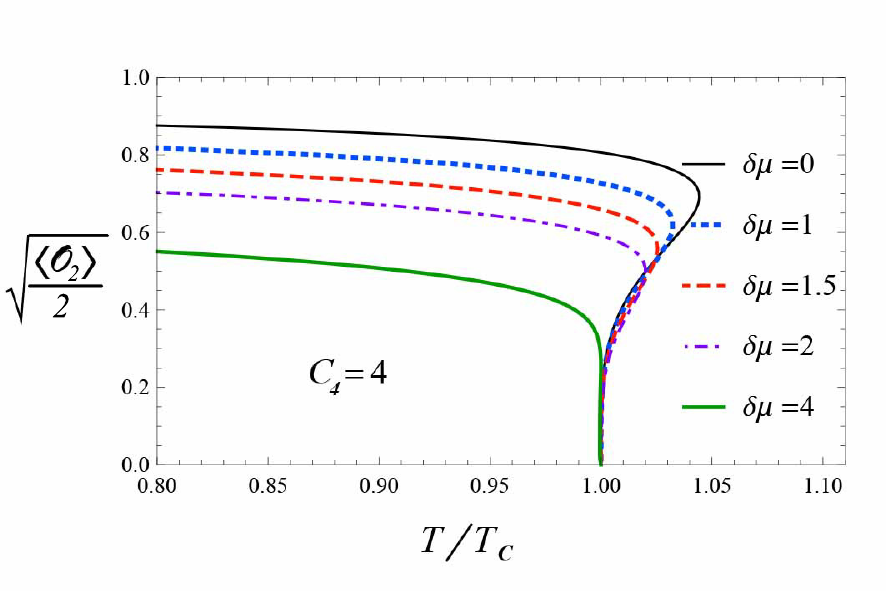}}        
       \qquad
        \subfloat[]{\includegraphics[width=0.5\columnwidth]{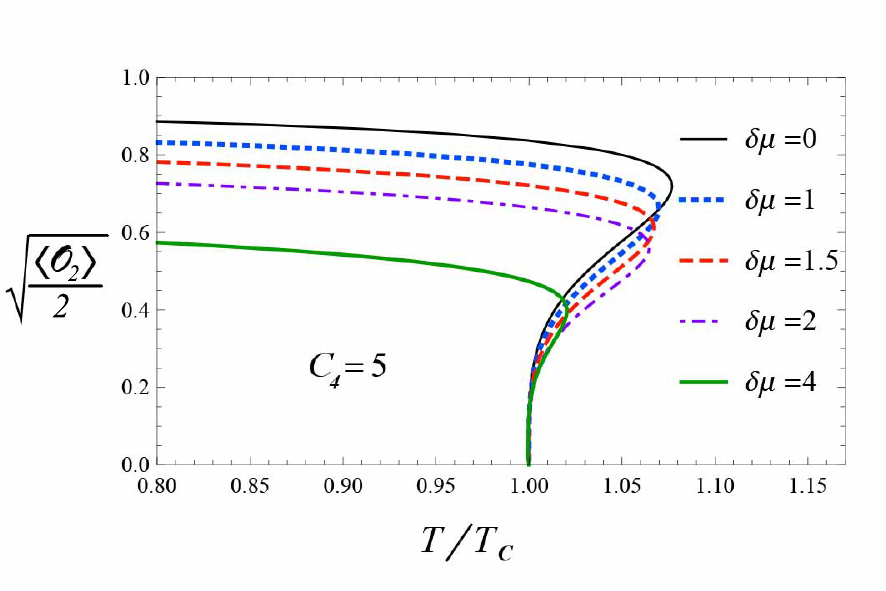}}
        \subfloat[]{\includegraphics[width=0.5\columnwidth]{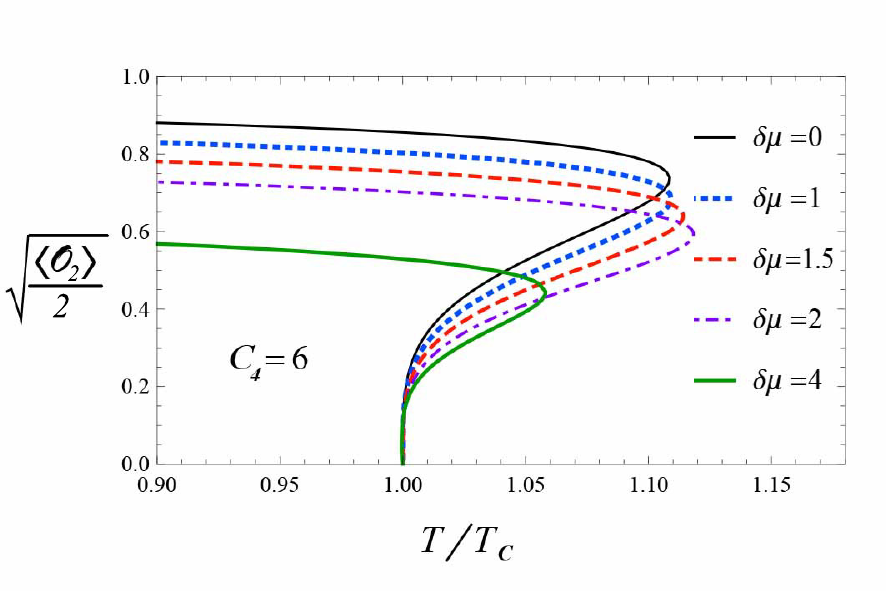}}
    \caption{Value of condensation as a function of temperature for function $\mathcal{F}(\psi)=\psi^{2}+C_{4}\psi^{4}$ with $C_4=0,2,3,4,5,6$ and $\delta \mu=0,0.5,1,1.5,2$.}
        \label{CSC4}
\end{figure}

\begin{figure}[h]
    \centering
        \subfloat[]{\includegraphics[width=0.5\columnwidth]{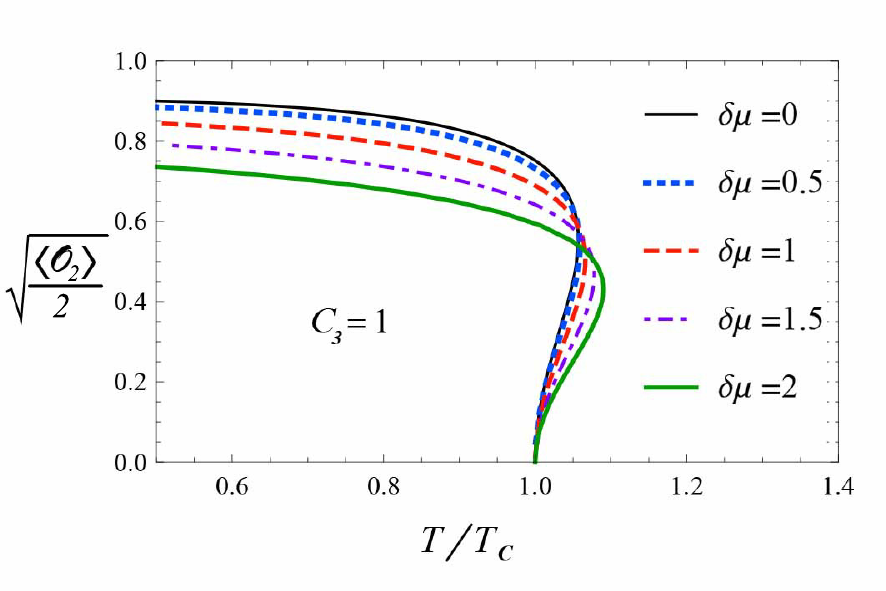}}
        \subfloat[]{\includegraphics[width=0.5\columnwidth]{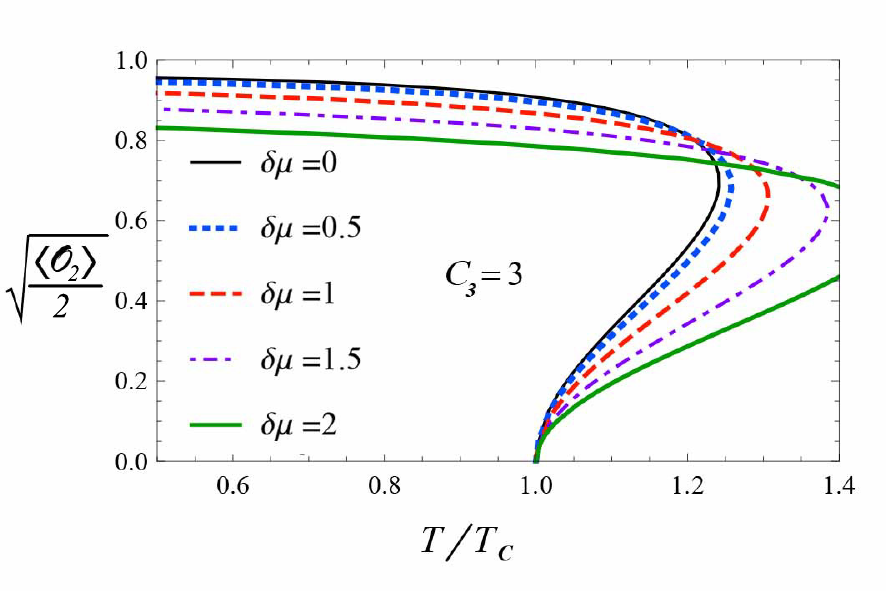}}
       \caption{The value of the condensate as a function of the temperature for function $\mathcal{F}(\psi)=\psi^{2}+C_{3}\psi^{3}$ with $C_3=1,3$ (left plot, right plot) and $\delta \mu=0,0.5,1,1.5,2$.}
        \label{CSC3}
\end{figure}
We start with the special case of $\mathcal{F}(\psi)=\psi^{2}+C_{4}\psi^{4}$ to identify the order of phase transitions in the interval  $0\leq\delta \mu/\mu\leq4$.
Figs. (\ref{CS2C4}) illustrates the change of phase transition order by increasing $C_4$.
Moreover, Fig. (\ref{CSC4}) demonstrates that the influence of reducing $C_4$ on phase transition is stronger in less unbalanced systems.
The results are detailed as follows:
\begin{itemize}
  \item Figs (\ref{CS2C4}) and (\ref{CSC4}) show the change of the phase transition order caused by increasing $C_4$. The second order phase transitions occurs for $0\leq C_{4}\lesssim 2$ and the first order ones occurs for $C_{4} \gtrsim 5$ (see Fig. (\ref{CSC4})). However, for the region $3\lesssim C_{4}\lesssim 4$ whether the phase transition is second or first order depends on the value of $\delta \mu/\mu$. The curves in Fig. (\ref{CS2C4}) (c) and (d) illustrate that our most unbalanced systems, i.e. $\delta \mu/\mu=2$ and $4$, do not undergo a first order phase transition even for $C_4=3$. As a result, increasing the imbalance in a system makes it harder to switch the order of phase transition from second to first by increasing $C_4$.
  \item We numerically check that the condensations approach zero as
  \begin{equation}\label{beta}
  \langle {\cal O}_2 \rangle \varpropto (T_c-T)^\beta,
  \end{equation}
  with mean field critical exponent $\beta = 1/2$ for the second order phase transitions. Thus, $\beta$ does not depend on neither $\delta \mu/\mu$ nor $C_4$.
\end{itemize}
%%%%%%%%%%%%%%%%%%
\subsubsection{The case of ${\cal F}(\psi)=\psi^2+C_3 \psi^3$}
As clearly shown in Fig. (\ref{CSC3}), in this case, first order phase transition occurs for any non-vanishing positive $C_3$. It is important to note that parameter $\delta \mu/\mu$ has no effect on the order of phase transition. Since all the phase transitions are first order, relation \eqref{beta} is not valid here.

%%%%%%%%%%%%%%%%%%
\subsubsection{ The case of ${\cal F}(\psi)=\psi^2-\psi^{\alpha}+ \psi^4$}

\begin{figure}[h]
    \centering    
        \subfloat[]{\includegraphics[width=0.5\columnwidth]{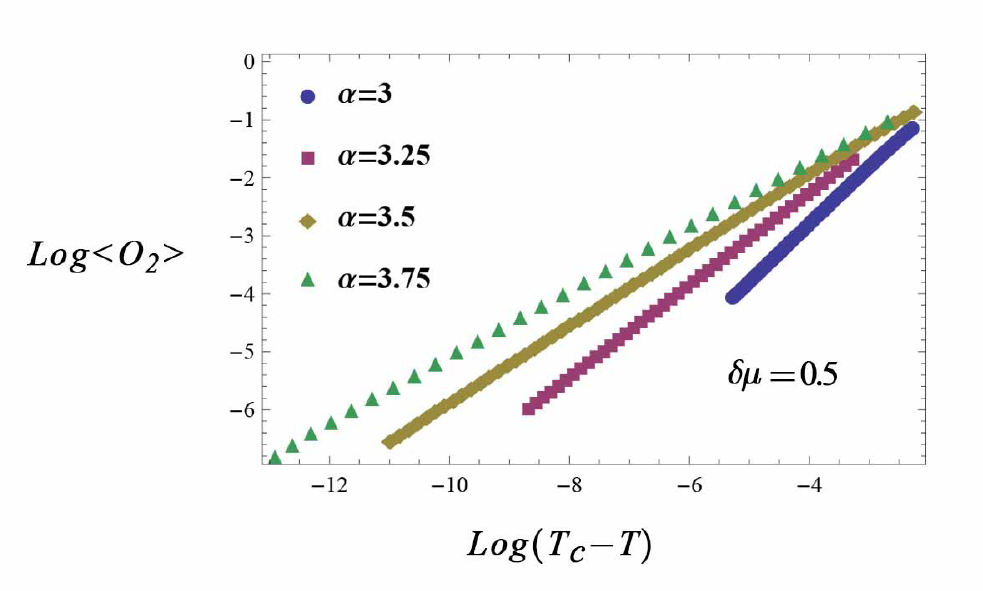}}
        \subfloat[]{\includegraphics[width=0.5\columnwidth]{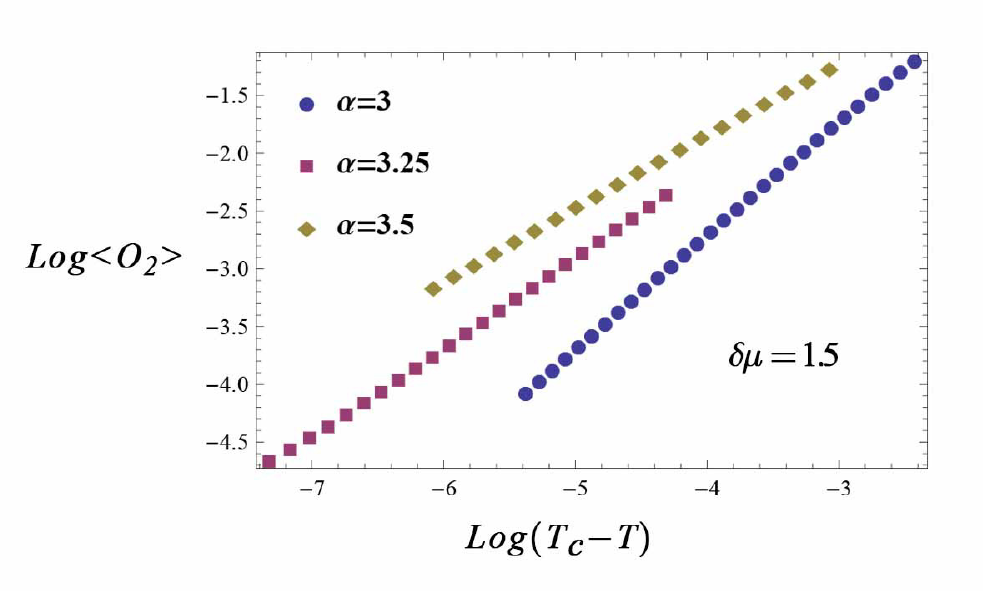}}        
       \qquad
        \subfloat[]{\includegraphics[width=0.5\columnwidth]{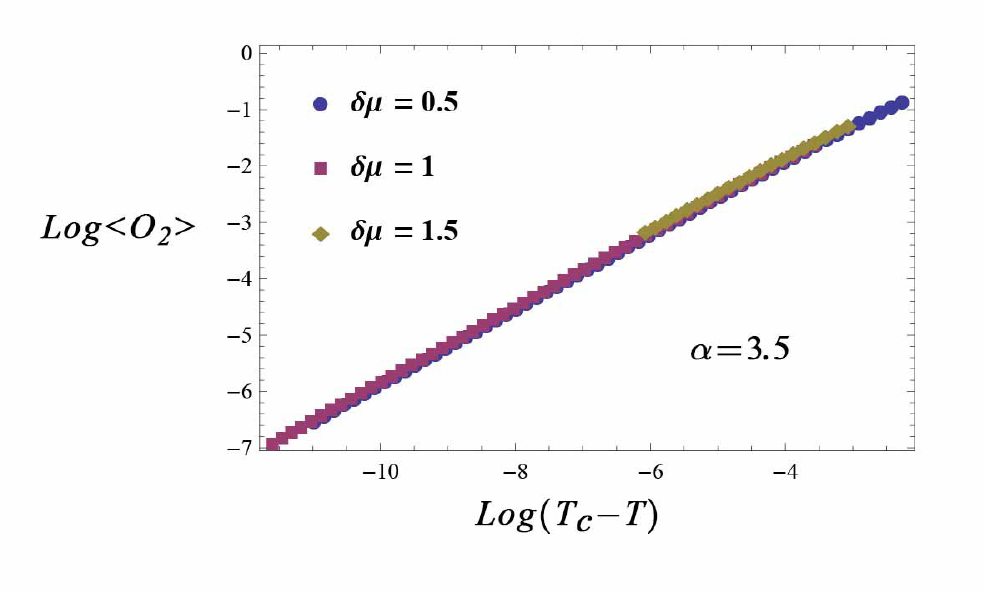}}
        \subfloat[]{\includegraphics[width=0.5\columnwidth]{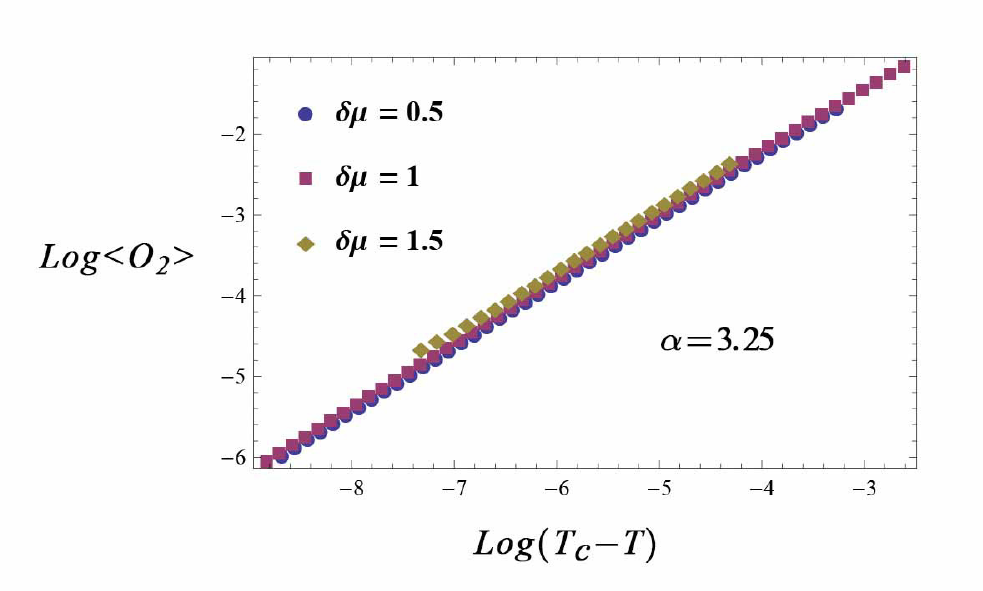}}
     \caption{The value of the condensate near the critical temperature for function ${\cal F}(\psi)=\psi^2-\psi^{\alpha}+ \psi^4$. Each plot in the first array indicates the condensation for fixed $\delta\mu=0.5,1.5$ (left plot, right plot) and various values of $\alpha$. The plots in the second array indicate condensation for fixed $\alpha=3.5,3.25$ (left plot, right plot) and different $\delta\mu=0.5,1,1.5$. It shows that imbalance does not violate relation~\eqref{beta2}.}
        \label{Beta}
\end{figure}

We are interested in investigating the effect of $\alpha$ (for $C_\alpha<0$) on critical exponent $\beta$ and searching for a non-mean field behavior.
We check that the relation
 \begin{equation}\label{beta2}
 \beta=(\alpha-2)^{-1},
  \end{equation}
from Ref. \cite{Franco:2009yz}, remains unchanged even in unbalanced systems. As indicated in  Fig. (\ref{Beta}), the above relation has been checked for a few different values of $\delta \mu/\mu$ when $3\leq\alpha<4$. The data in Fig. (\ref{Beta}) (c) and (d) show that the imbalance clearly has nothing to do with the gradient of condensation plot near the critical temperature.
It is worth to mention that in the relation \eqref{beta2}, $\beta$ is larger than the mean field critical exponent for $3\leq\alpha<4$. Such behavior causes the suppression of the fluctuations and the stability of the condensation as observed in the Gross-Neveu model for massless fermions \cite{Rosa:2000ju}. Moreover, it likely indicates the existence of long-range interaction and chiral symmetry in the boundary theory \cite{Rosenstein:1993zf,SilvaNeto:1998db}.

%%%%%%%%%%%%%%%%%%%%%%%%%%%%%%%%%%%%%%
%%%%%%%%%%%%%%%%%%%%%%%%%%%%%%%%%%%%%%
\section{Conductivity}\label{sec3}

In this section, we study the conductivity properties of our model. In addition to considering mixed spin-electric linear response to the external gauge fields fluctuations, here we add the thermal effects, namely the thermo-electric and thermo-spin linear response to the temperature gradient. Therefore, one can define the conductivity matrix as follows:
\begin{eqnarray}\label{CtMatrix}
\begin{pmatrix}  J^A \\ Q \\ J^B \end{pmatrix} = \begin{pmatrix}  \sigma_A & \alpha T & \gamma \\ \alpha T &
\kappa T & \beta T \\ \gamma & \beta T & \sigma_B \end{pmatrix} \cdot \begin{pmatrix}  E^A \\ -\frac{\nabla T}{T} \\ E^B \end{pmatrix} \,,
\end{eqnarray}
which encodes the whole system response. The diagonal components $\sigma_A$, $\sigma_B$, and $\kappa T$ stand for ``electric'', ``spin'', and ``thermal'' conductivities, respectively. Furthermore, the off-diagonal components indicate mixed effects; i.e. $\gamma$, $\alpha T$, and $\beta T$ indicate the ``mixed'', ``thermo-electric'', and ``thermo-spin'' response, respectively. The symmetry of this matrix is a result of time-reversal invariance \cite{Son:1987,Johnson:1987,Hartnoll:2008kx} .

To study the transport behavior of our system, we take a small variation of the sources and the consequent current flows.
More specifically, to calculate conductivities in the boundary field theory side, we need to turn on the perturbation of the gauge fields $A$ and $B$ in the direction $x$ with time dependent function $e^{-i\omega t}$ in the bulk. Afterwards, by substituting Einstein equation in the two Maxwell equations on the background, and eliminating metric fluctuations, one arrives at the two following linear differential equations:
\begin{equation}\label{eqA}
 A_x'' + \left(\frac{g'}{g} - \frac{\chi'}{2}\right) A_x'
 + \left( \frac{\omega^2}{g^2}e^{\chi} - \frac{2 q^2 \mathcal{F}(\psi)}{g} \right) A_x
 - \frac{\phi'}{g} e^{\chi} \left(B_x v' + A_x \phi' \right) = 0\, ,
\end{equation}
\begin{equation}\label{eqB}
 B_x'' + \left(\frac{g'}{g} - \frac{\chi'}{2}\right) B_x'
 + \frac{\omega^2}{g^2}e^{\chi} B_x
 - \frac{v'}{g} e^{\chi} \left(B_x v' + A_x \phi' \right) = 0\, .
\end{equation}
Note that the backreaction leads to coupled differential equations. This event is responsible for appearing the mixed spin-electric transport properties in a system \cite{Bigazzi:2011ak}.
We can consider near-horizon behavior ansatz 
\begin{eqnarray}
 & A_x(r) & = \left(1-\frac{r_H}{r}\right)^{i a \omega} \left[ 1 + a_1 \left(1-\frac{r_H}{r}\right) + ... \right]\, ,\\ \label{asintoticiB}
 & B_x(r) & = \left(1-\frac{r_H}{r}\right)^{i a \omega} \left[ 1 + b_1 \left(1-\frac{r_H}{r}\right) + ... \right]\,,
\end{eqnarray}
which also impose ingoing boundary conditions at horizon. In addition, the asymptotic behavior of fields around boundary $r \to \infty$ is
\begin{eqnarray}\label{boufie}
 A_x(r) &= A_x^{(0)} + \frac{1}{r} A_x^{(1)} + ... \, ,\\
 B_x(r) &= B_x^{(0)} + \frac{1}{r} B_x^{(1)} + ... \, ,\\
 g_{tx}(r) &= r^2 g_{tx}^{(0)} - \frac{1}{r} g_{tx}^{(1)} + ...\, .
\end{eqnarray}
Using introduced method in \cite{Bigazzi:2011ak}, we can finally get
\begin{eqnarray}\label{Ct}
\sigma_A = - \frac{i}{\omega} \frac{A_x^{(1)}}{A_x^{(0)}}|_{g_{tx}^{(0)}=B_x^{(0)}=0}\ , \nonumber \\
\gamma = - \frac{i}{\omega} \frac{B_x^{(1)}}{A_x^{(0)}}|_{g_{tx}^{(0)}=B_x^{(0)}=0}\ \\
=- \frac{i}{\omega} \frac{A_x^{(1)}}{B_x^{(0)}}|_{g_{tx}^{(0)}=A_x^{(0)}=0}\ , \nonumber \\
\sigma_B  = - \frac{i}{\omega} \frac{B_x^{(1)}}{B_x^{(0)}}|_{g_{tx}^{(0)}=A_x^{(0)}=0}\ . \nonumber
\end{eqnarray}
The thermo-electric and the thermo-spin conductivities are also obtained as follows:
\begin{eqnarray}\label{TCt}
\alpha T =  \frac{Q}{E^{A}}|_{g_{tx}^{(0)}=B^{(0)}=0} = \frac{i \rho}{\omega} - \mu \sigma_A -\delta\mu \gamma \, ,\\
\beta T =  \frac{Q}{E^{B}}|_{g_{tx}^{(0)}=A_x^{(0)}=0} = \frac{i \delta \rho}{\omega} - \delta \mu \sigma_B -\mu \gamma\, . \nonumber
\end{eqnarray}
Finally, one can find that the non-canonical thermal conductivity is given by
\begin{equation}\label{kappa}
\kappa\,T=\frac{i}{\omega}[\epsilon + p -2\mu\rho -2\delta\mu\delta\rho] + \sigma_A \mu^2 + \sigma_B
\delta\mu^2 + 2 \gamma \mu \delta\mu\,,
\end{equation}
where we have considered pressure $p=\epsilon/2$, like its value in \cite{Bigazzi:2011ak}, in order to account for contact terms not directly implemented by the previous computations (see Herzog's review in \cite{Herzog:2009xv}). To find more details about equations \eqref{Ct}, \eqref{TCt}, and \eqref{kappa} see \cite{Bigazzi:2011ak}.
By numerically solving equations \eqref{eqA} and \eqref{eqB} and utilizing \eqref{Ct}, \eqref{TCt}, and \eqref{kappa} we are able to study the effects of the model parameters and the imbalance on all the conductivity types.

%%%%%%%%%%%%%%%%%%%%%%%%%%%%%%%%%%%%%%%

\begin{figure}[h]
    \centering
        \subfloat[]{\includegraphics[width=0.5\columnwidth]{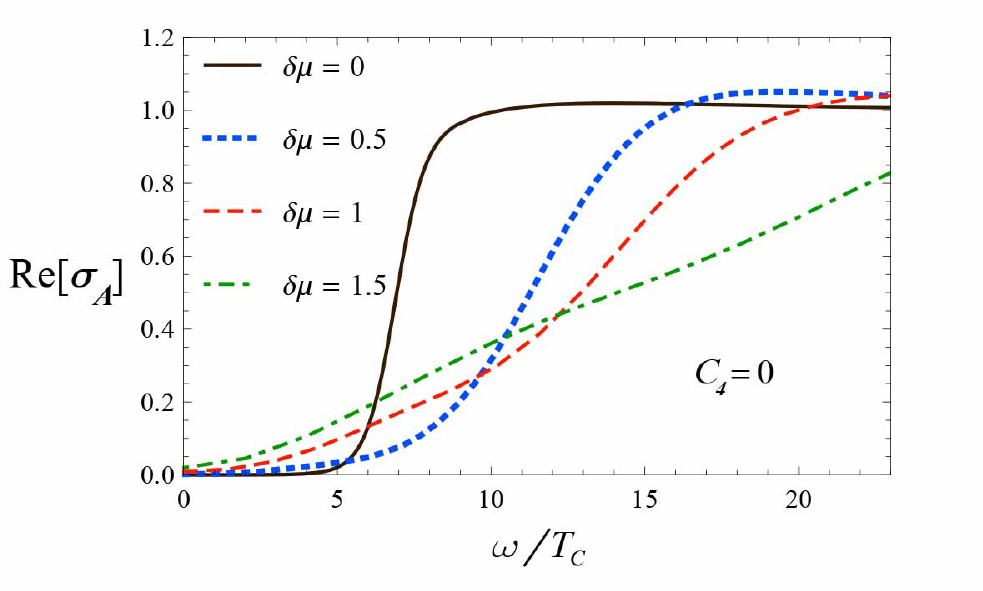}}
        \subfloat[]{\includegraphics[width=0.5\columnwidth]{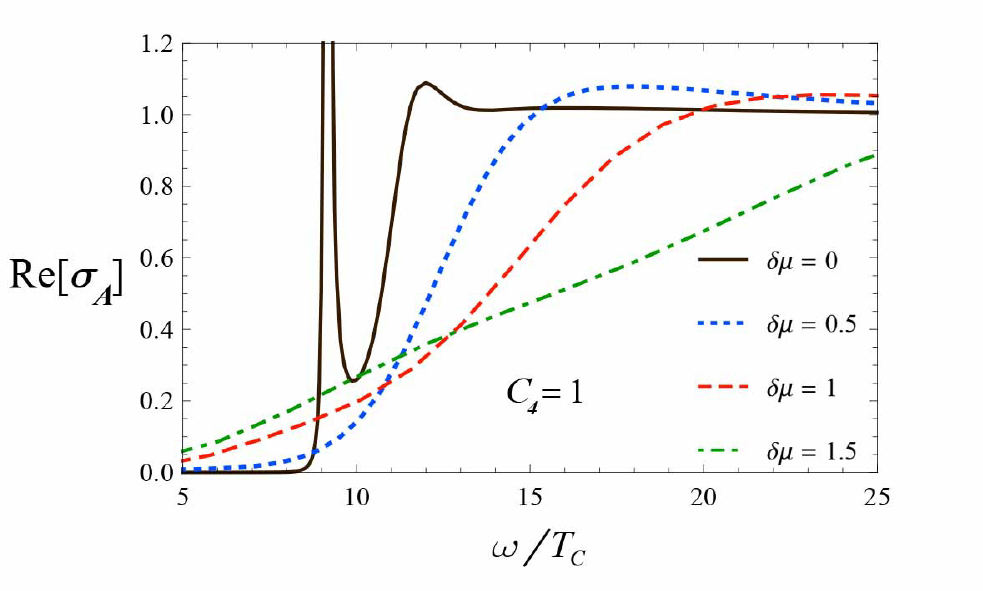}}
      \qquad
        \subfloat[]{\includegraphics[width=0.5\columnwidth]{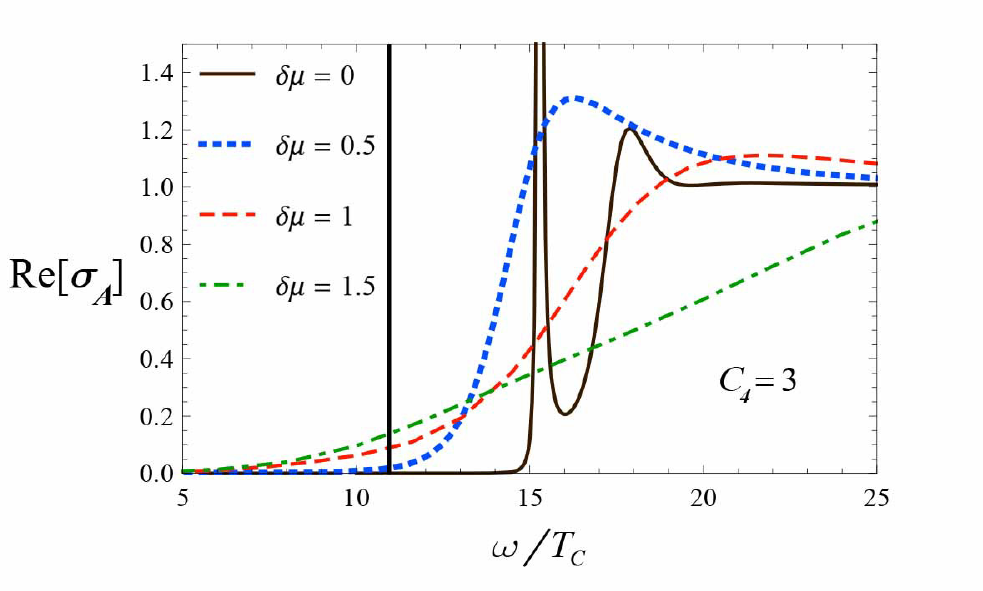}}
        \subfloat[]{\includegraphics[width=0.5\columnwidth]{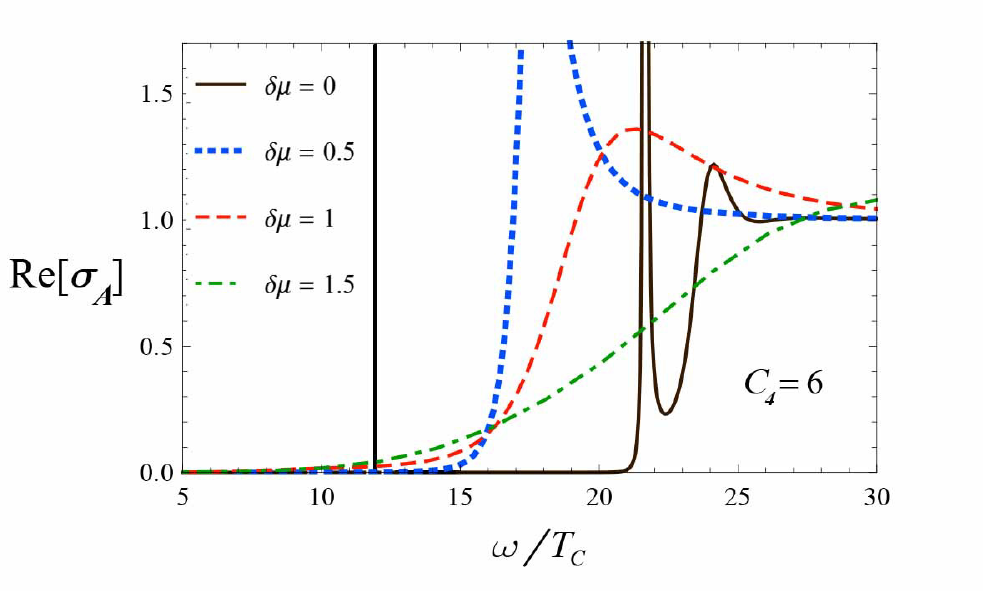}}
      \qquad
        \subfloat[]{\includegraphics[width=0.5\columnwidth]{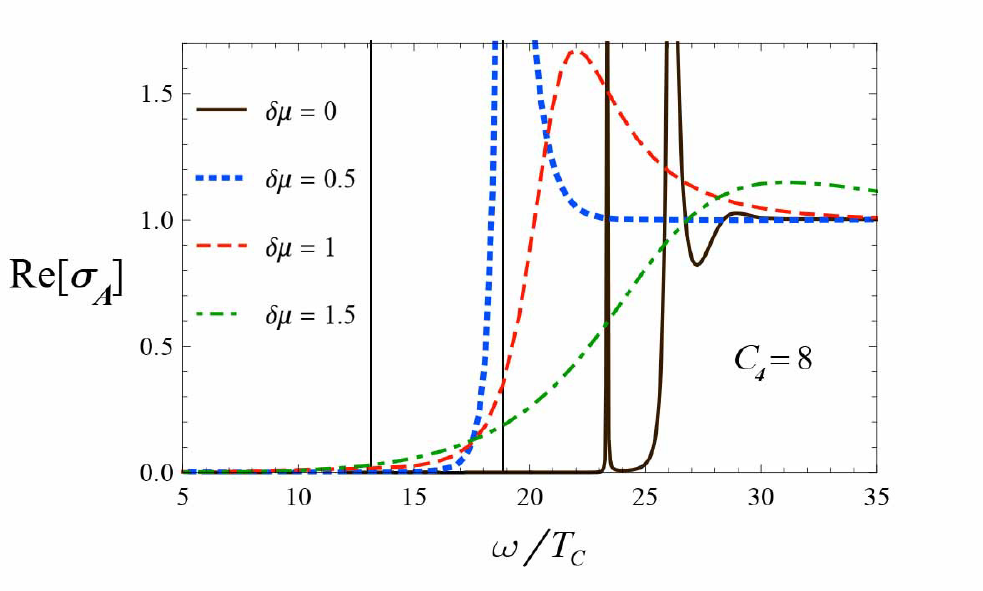}}
        \caption{The optical electric conductivity in terms of $\omega/T_c$ for function ${\cal F}(\psi)=\psi^2+C_4\psi^4$ and $\delta\mu=0,0.5,1,1.5$ (solid curve, dotted curve, dashed curve, and dot-dashed curve). In each figure we have fixed values $C_4=0,1,2,3,6,8$ for figures (a), (b), (c), (d), and (e), respectively.}
        \label{CTsAC4}
\end{figure}

\begin{figure}[h]
    \centering
        \includegraphics[width=0.5\columnwidth]{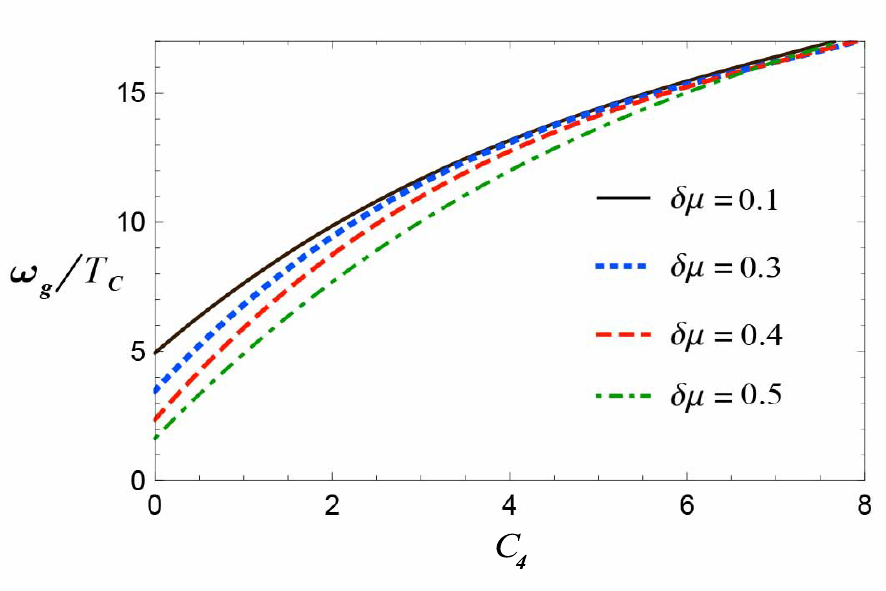}
          \caption{Plot of the $\omega_g/T_c$ as a function of $C_4$ for $\delta\mu=0.1,0.3,0.4,0.5$ (solid curve, dotted curve, dashed curve, dot-dashed curve). Here, we have fixed $T=0.3Tc$ and considered the numerical threshold $Re[\sigma]=0.005$ to numerically define $\omega_g$.}
        \label{gap}
\end{figure}
\subsection{Diagrams and behaviors}

We restrict ourselves to the case where the temperature takes value $T=0.3T_c$. As before, we assume the chemical potential equals to one ($\mu=1$) over this section. 
Note that the imaginary part of the conductivity has a pole at $\omega=0$, which translates in a delta function at the same point in the real part, according to the Kramers-Kroning relation.
Since we are working with the fully backreacted solution, translational invariance is preserved due to the lack of dissipation in probe approximation. 
Because of the Ferrell-Glover-Tinkham sum rule, the area under the curves must be constant at different temperatures. Therefore, we have a depletion at small frequencies to compensate the development of the delta function at $\omega=0$ \cite{Hartnoll:2008kx}.
According to the terminology used in \cite{Bigazzi:2011ak}, we take the ``pseudo-gap'' idiom to describe the depletion at small frequencies; since the real part of the electric conductivity appears exponentially small with respect to $T$, this is not exactly zero even at $T=0$. 
 
 %%%%%%%%%%%%%%%%%%%%%%%%%
\subsubsection{Conductivity Behavior with respect to the variation of $\delta \mu/ \mu$}

\begin{figure}[h]
\centering
  \subfloat[]{\includegraphics[width=0.5\columnwidth]{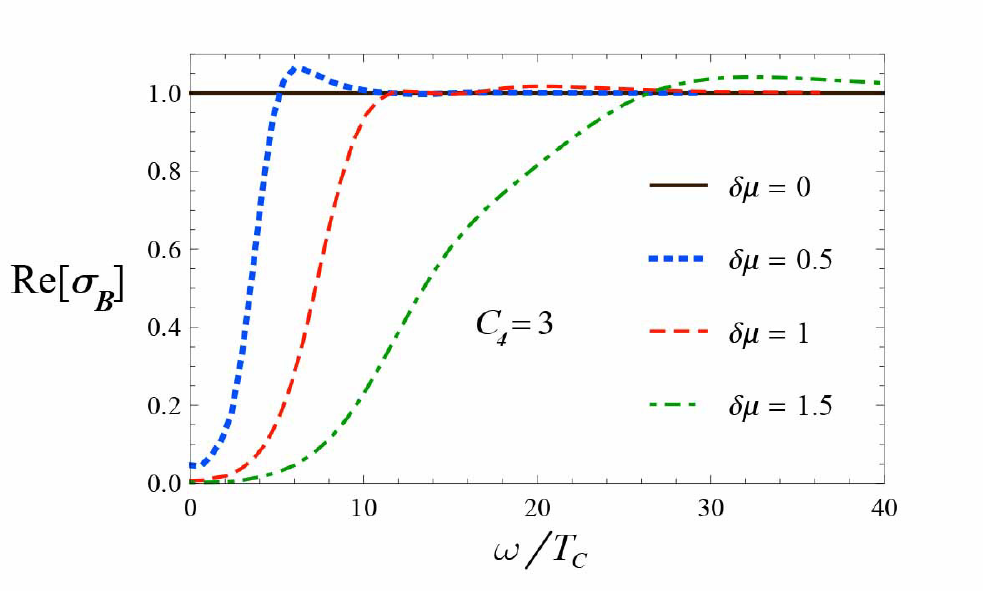}}
   \subfloat[]{\includegraphics[width=0.5\columnwidth]{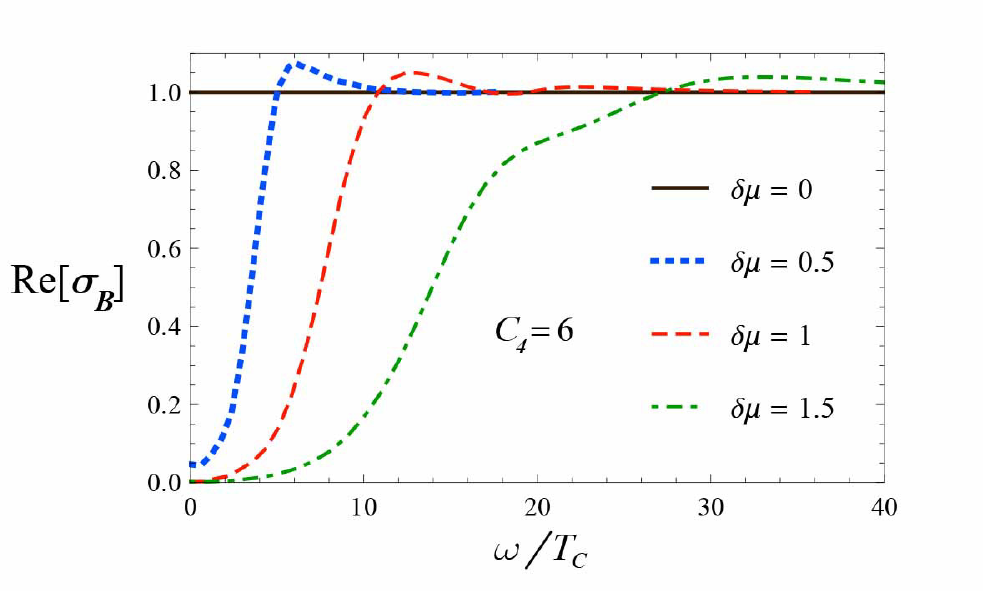}}
      \caption{The real part of spin conductivity in terms of $\omega/T_c$ for $\delta\mu=0,0.5,1,1.5$ (solid curve, dotted curve, dashed curve, dot-dashed curve) and function ${\cal F}(\psi)=\psi^2+C_4\psi^4$. We have considered the non-vanishing $C_4=3$ (left) and  $C_4=6$ (right) in order to highlight the fluctuations.}
\label{CTsBdm}
\end{figure}

\begin{figure}[h]
    \centering
        \subfloat[]{\includegraphics[width=0.5\columnwidth]{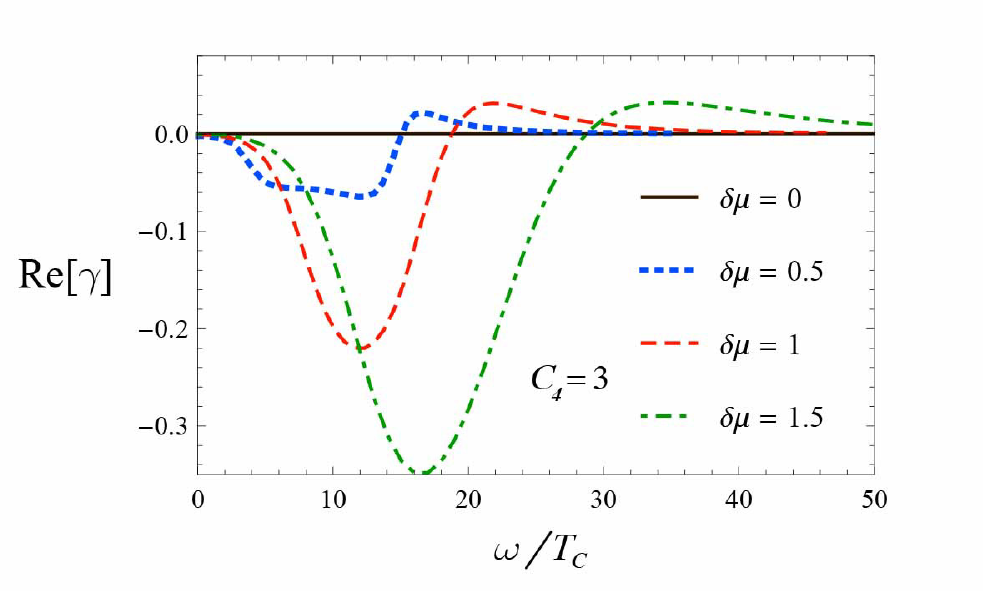}}
        \subfloat[]{\includegraphics[width=0.5\columnwidth]{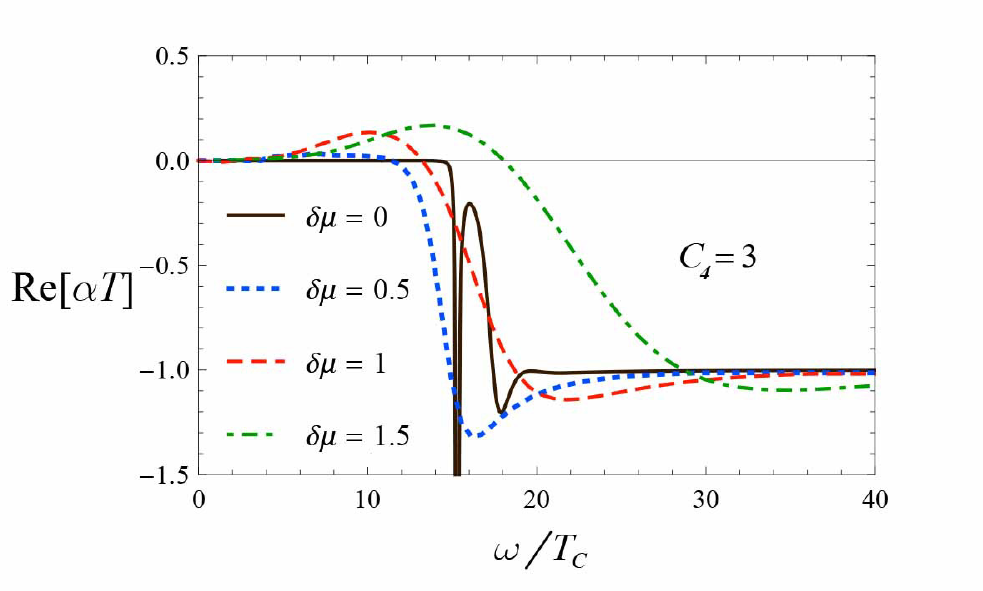}}
      \qquad
       \subfloat[]{\includegraphics[width=0.5\columnwidth]{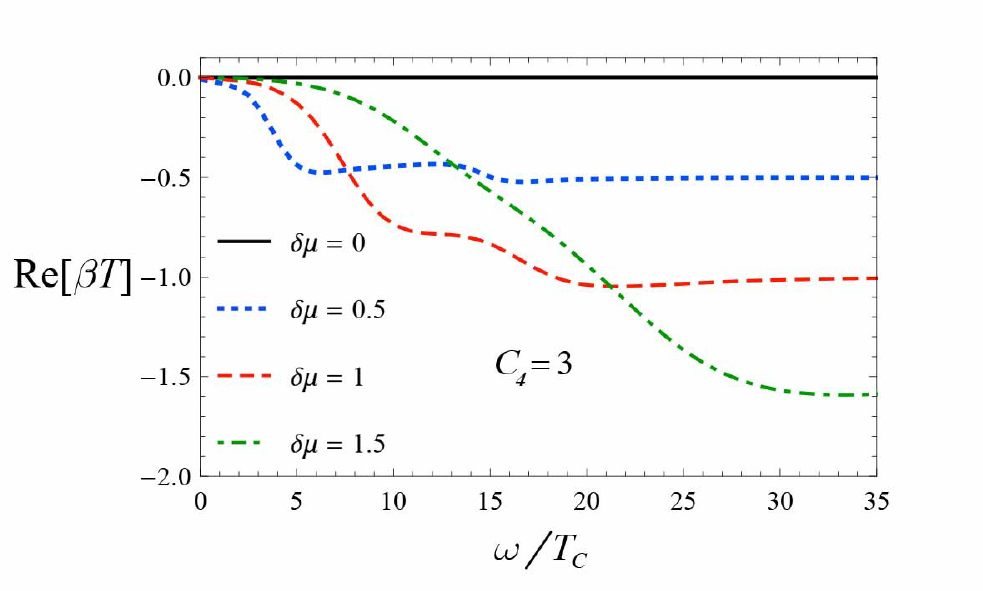}}
        \subfloat[]{\includegraphics[width=0.5\columnwidth]{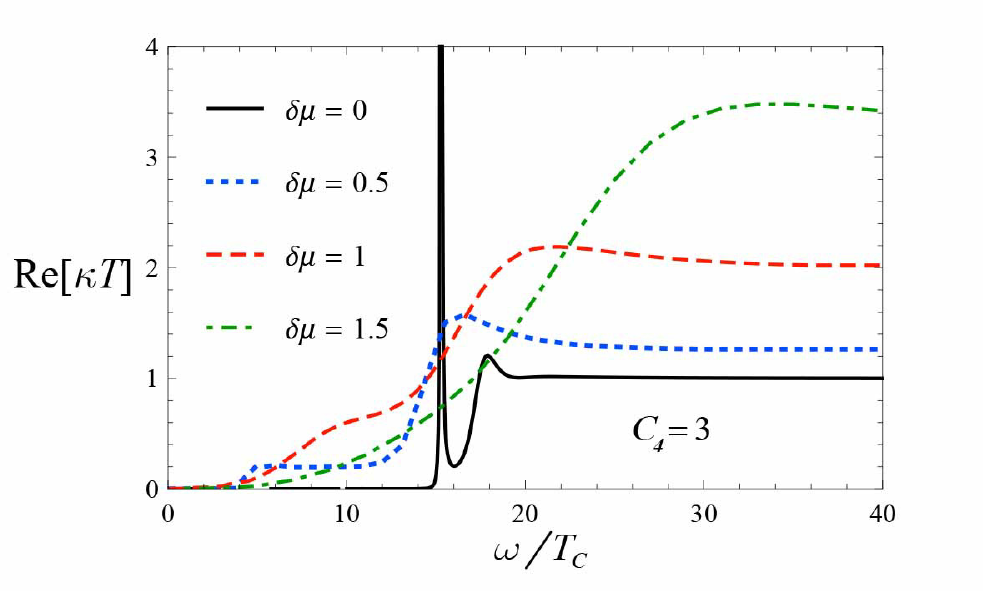}}
      \caption{The real part of mixed, thermo-electric, thermo-spin, and thermal conductivities (figures (a), (b), (c), and (d)) in terms of $\omega/T_c$ for  $\delta\mu=0,0.5,1,1.5$ (solid curve, dotted curve, dashed curve, dot-dashed curve) and function ${\cal F}(\psi)=\psi^2+3\psi^4$. We have considered the non-vanishing $C_4$ to highlight the fluctuations.}
 \label{CTdm}
\end{figure}

We first carry out analysis on the conductivities behavior in the presence of an imbalance. We are actually looking for the validity of the results achieved in \cite{Bigazzi:2011ak} for our model. We can therefore consider a fixed form of function ${\cal{F}}(\psi)$ to investigate the conductivities for various imbalances, i.e. $\delta\mu/\mu=0,0.5,1,1.5$.

Fig. (\ref{CTsAC4}) illustrates the decline in the pseudo-gap of electric conductivity as a system becomes more and more unbalanced, which is also reported in \cite{Bigazzi:2011ak}. For example, Fig. (\ref{CTsAC4}) (a) demonstrates that the pseudo-gap of the system with $\delta\mu/\mu=1.5$ almost vanishes. Moreover, one can easily see that the difference between the pseudo-gap values for different imbalances becomes negligible at a high enough value of $C_4$. 
In order to clarify this point, we depict $\omega_g/T_{c}$ as a function of $C_4$ for different values of $\delta\mu/\mu$ in Fig (\ref{gap}).
One can realize that the differences between $\omega_g/T_{c}$ of various unbalanced systems almost vanishe for large $C_4$s.

Fig. (\ref{CTsAC4}) also indicates that the increasing coefficient parameter $C_4$ may make the coherent peak turn to a delta function (See Fig. (\ref{CTsAC4}) and (\ref{CTsA}) (a)).
It should be noted that this is followed by the change of $\omega_g$ position from the frequency of the delta function to near the frequency of the next peak. Therefore, $\omega_g/T_{c}$ of various unbalanced systems do not converge to the same value for large amounts of $C_4$, but by ignoring these jumps in pseudo-gap we can see that all values of $\omega_g/T_{c}$ approach to the same amount.

In addition, imbalance disturbs the constant values of the spin and mixed conductivities. Note that optical spin and mixed conductivities of a balanced system are constant values of $1$ and $0$, respectively. In our model, the optical spin and mixed conductivities of unbalanced systems relaxe to these values at large $\omega$ after some fluctuations.

It is obvious from Fig. (\ref{CTsBdm}) that the optical spin conductivity becomes more and more depleted at small frequencies by growing the imbalance in contrast to the electric conductivity \cite{Bigazzi:2011ak}.
This opposite behavior of the electric and spin conductivities with respect to increasing $\delta\mu / \mu$ is usually interpreted as a separation of the dynamics of charge and spin degrees of freedom  \cite{Bellazzini:2008mn,Bigazzi:2011ak}.

The real part of the mixed conductivity is depicted in Fig. (\ref{CTdm}) (a). It shows a number of fluctuations for unbalanced systems. One can see that not only does the increase of the imbalance intensify these fluctuations, but it also shifts them to larger frequencies. 

The real part of the thermo-electric conductivity for function ${\cal{F}}(\psi)=\psi^2+3\psi^4$ are represented in Fig. (\ref{CTdm}) (c).
They show some fluctuations of the conductivity before converging to $-1$ at larger frequencies. More unbalanced systems (systems in the range of $\delta\mu / \mu=1$ and $1.5$) also tend to generate a positive peak in the conductivity at lower frequencies. This behavior, therefore, kills the pseudo-gap in such systems. 
In the next subsection, we also show that the increase of $C_4$ not only does not disturb the general behavior with imbalance but also amplifies it.

Imbalance also turns on the thermo-spin conductivity. Fig. (\ref{CTdm}) (c) shows that more unbalanced systems possess larger negative conductivities.

In the case of thermal conductivity, it worth mentioning that there are no obvious differences between the pseudo-gaps width of unbalanced cases. Fig. (\ref{CTdm}) (d) illustrates that the only pseudo-gap which is comparably different from the others belongs to the balanced system. In fact, the conductivity pseudo-gap is wider for the balanced system.
Therefore, it seems that there is a non-monotonic behavior with imbalance of the thermal conductivity in the small frequency region. This is similar to the behavior of unbalanced holographic superconductors built upon Higgs mechanism in \cite{Bigazzi:2011ak}.
It should also be noted that the real part of the thermal conductivity of systems with different imbalances does not rest to a same value at large frequencies, like the thermo-spin ones.

%%%%%%%%%%%%%%%%%%%%%%%%%%%%%%%%%%%%%%%%%%%%%%%%%%%%%%%%%
\subsubsection{Conductivity Behavior with respect to the variation of $C_4$}

\begin{figure}[t]
    \centering
        \subfloat[]{\includegraphics[width=0.5\columnwidth]{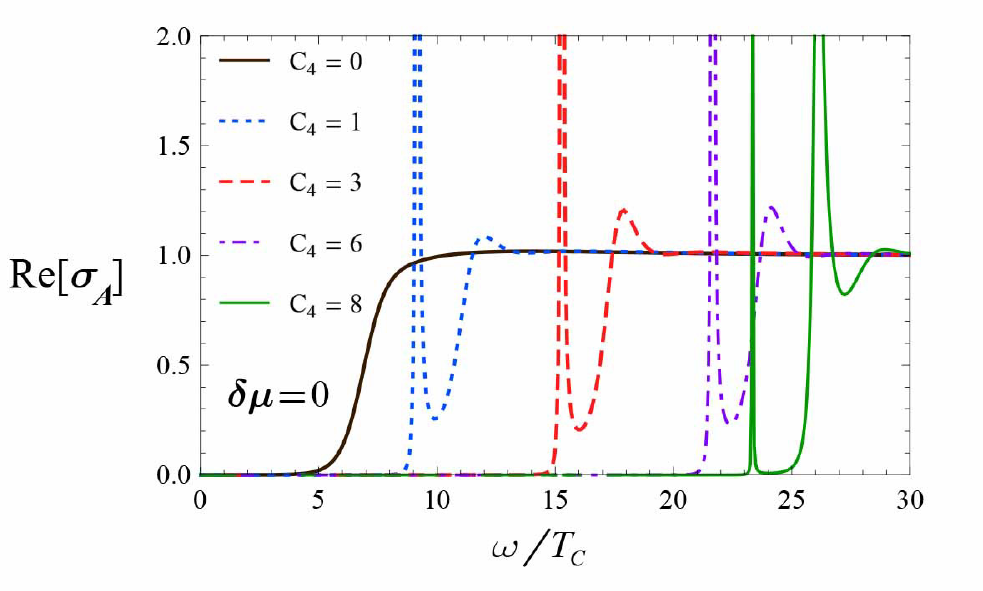}}
        \subfloat[]{\includegraphics[width=0.5\columnwidth]{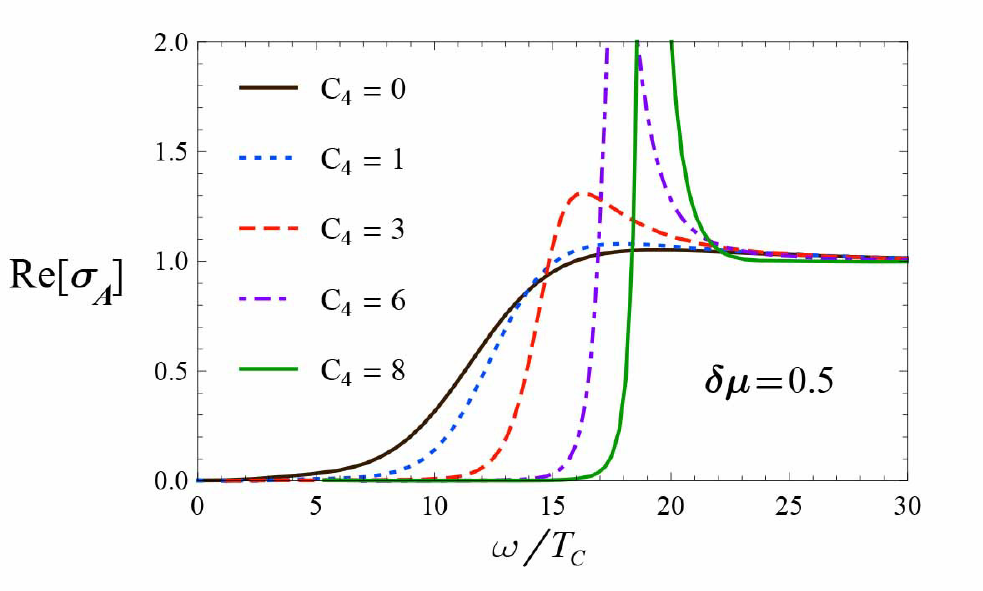}}
      \qquad
        \subfloat[]{\includegraphics[width=0.5\columnwidth]{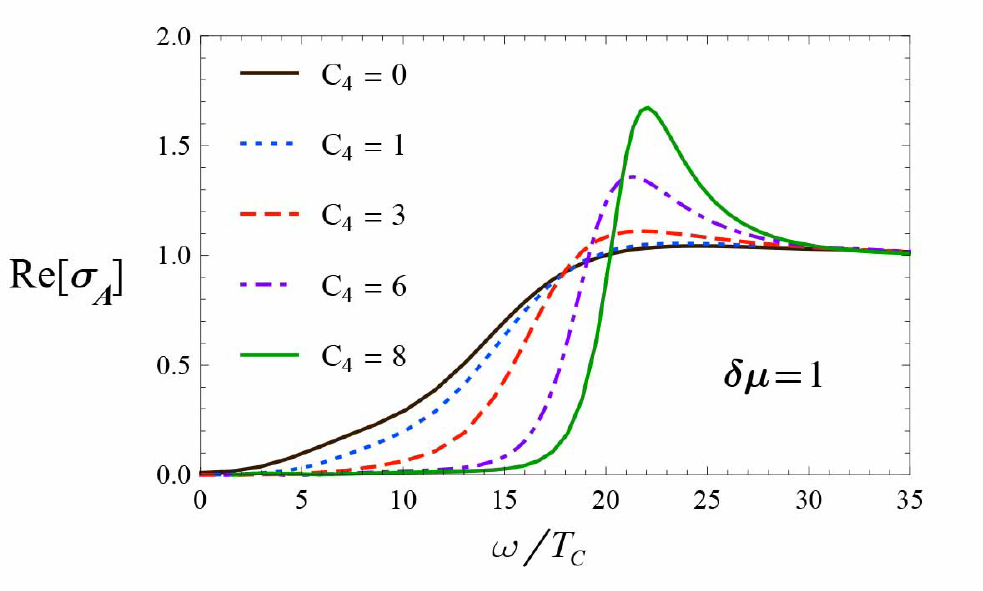}}
        \subfloat[]{\includegraphics[width=0.5\columnwidth]{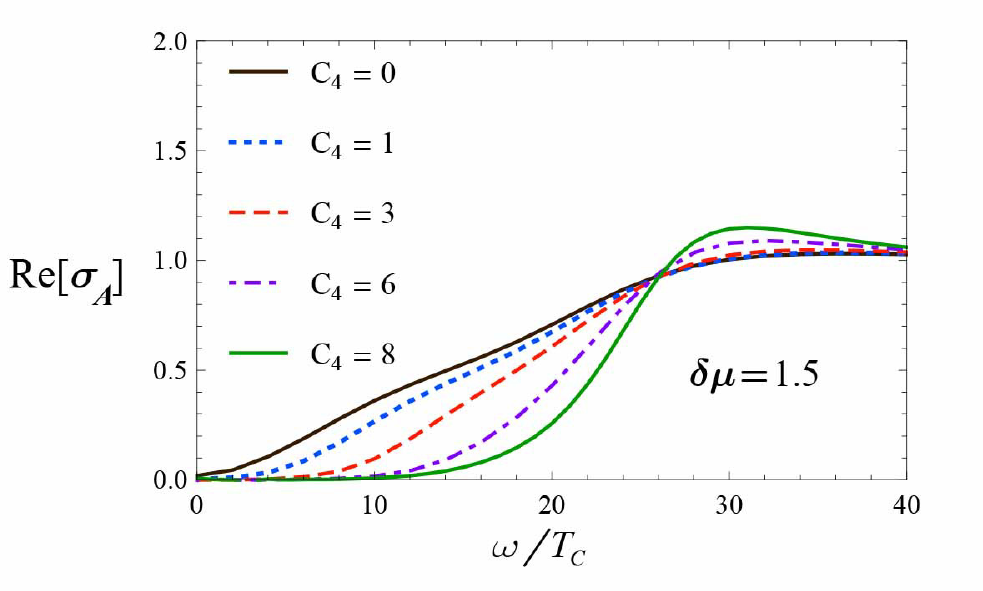}}
        \caption{The real part of the mixed conductivity in terms of $\omega/T_c$ for function ${\cal F}(\psi)=\psi^2+C_4\psi^4$ with $C_4=0,1,3,6,8$ (solid curve, dotted curve, dashed curve, dot-dashed curve, and pale (green) solid curve) and $\delta\mu=0,0.5,1,1.5$ (figures (a), (b), (c), and (e)).}
    \label{CTsA}         
\end{figure}

\begin{figure}[h]
    \centering
        \subfloat[]{\includegraphics[width=0.5\columnwidth]{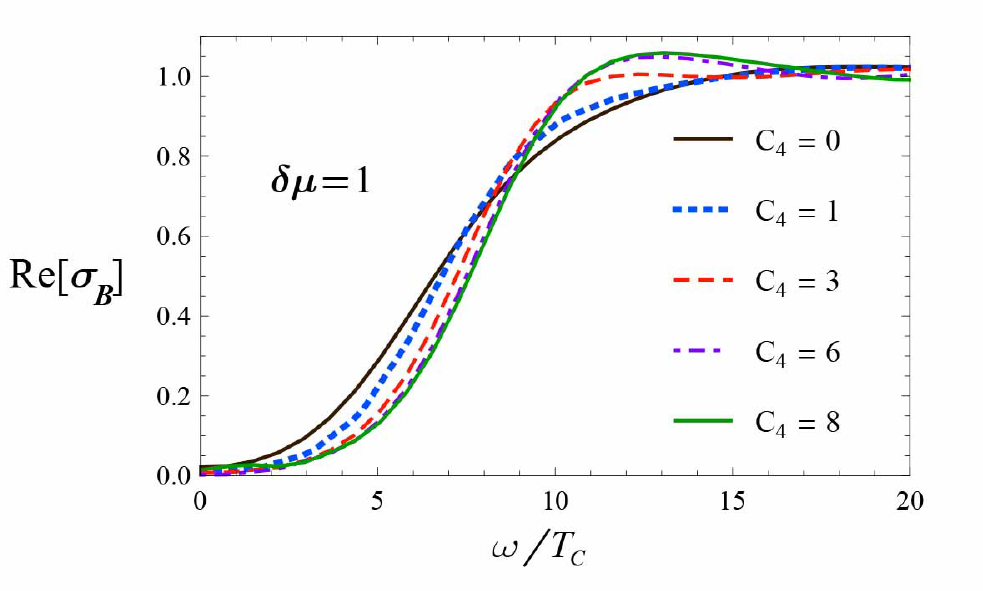}}
        \subfloat[]{\includegraphics[width=0.5\columnwidth]{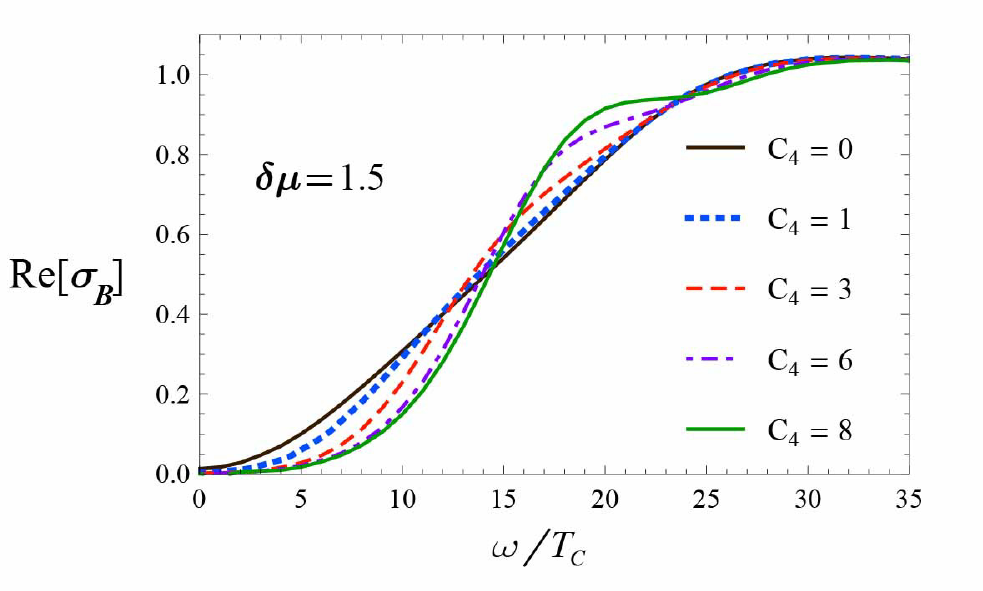}}
           \caption{The real part of the spin conductivity in terms of $\omega/T_c$ for function ${\cal F}(\psi)=\psi^2+C_4\psi^4$ with $C_4=0,1,3,6,8$ (solid curve, dotted curve, dashed curve, dot-dashed curve, and pale (green) solid curve). The conductivity of systems with $\delta\mu=1$ and $1.5$ are presented in the left and the right figure respectively.}
            \label{CTsB} 
\end{figure}

\begin{figure}[h]
    \centering
       \subfloat[]{\includegraphics[width=0.5\columnwidth]{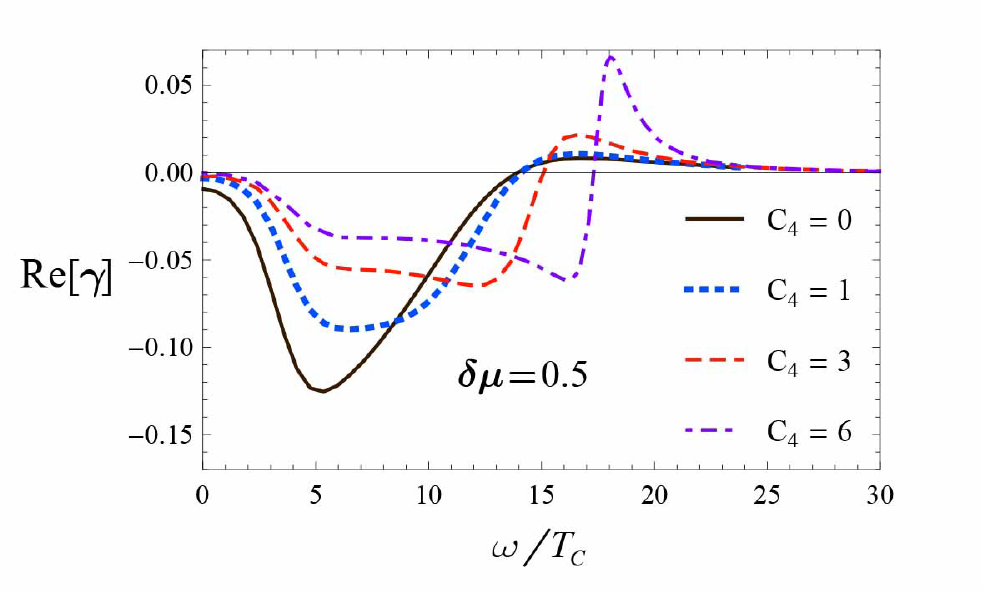}}
       \subfloat[]{\includegraphics[width=0.5\columnwidth]{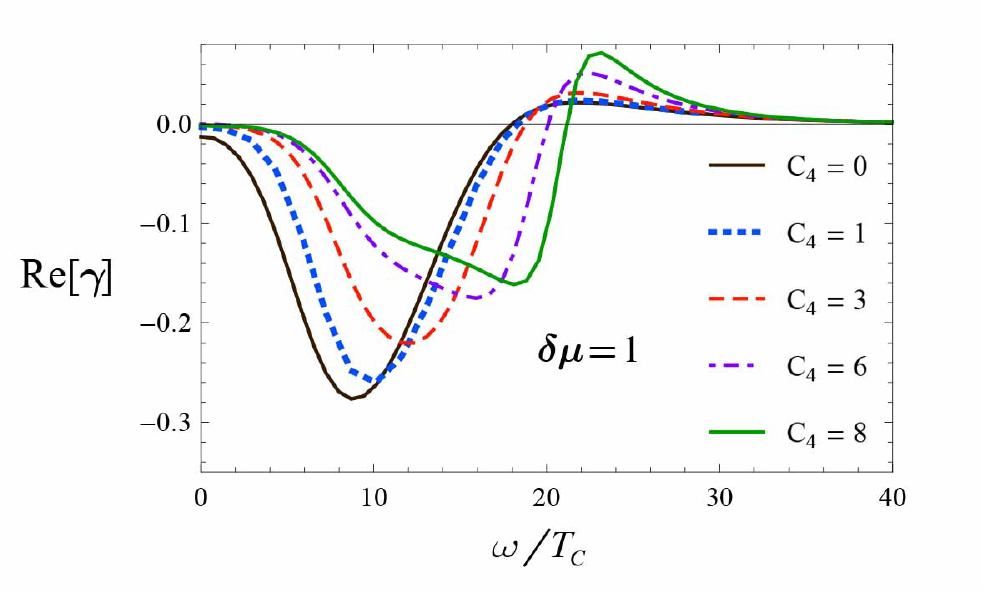}}
     \qquad
       \subfloat[]{\includegraphics[width=0.5\columnwidth]{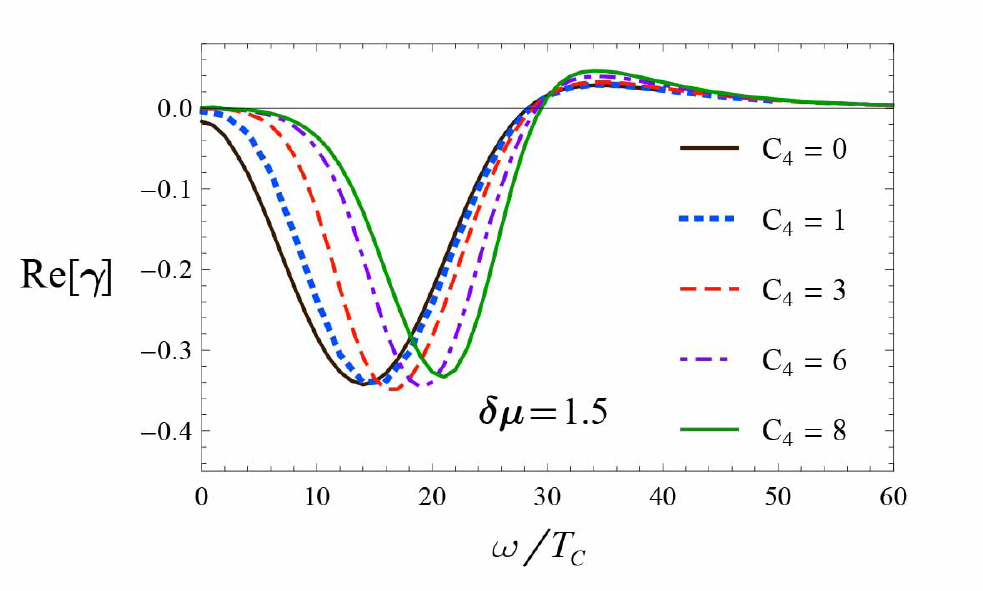}}
          \caption{The real part of the mixed conductivity in terms of $\omega/T_c$ for function ${\cal F}(\psi)=\psi^2+C_4\psi^4$ with $C_4=0,1,3,6,8$ (solid curve, dotted curve, dashed curve, dot-dashed curve, and pale (green) solid curve) and $\delta\mu=0.5,1,1.5$ (figures (a), (b), and (c)).}
  \label{CTg} 
\end{figure}

\begin{figure}[h]
    \centering
        \subfloat[]{\includegraphics[width=0.5\columnwidth]{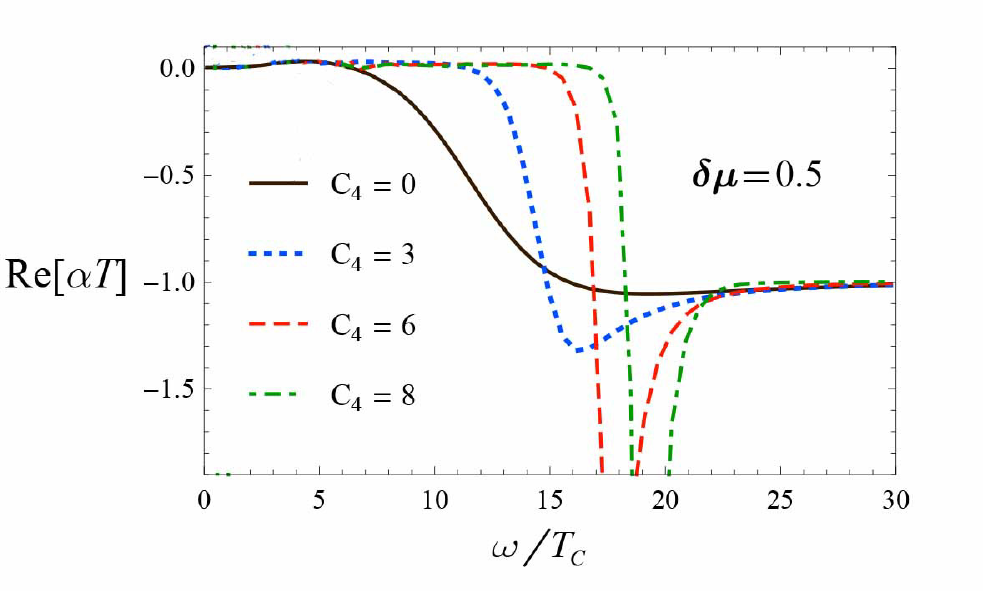}}
        \subfloat[]{\includegraphics[width=0.5\columnwidth]{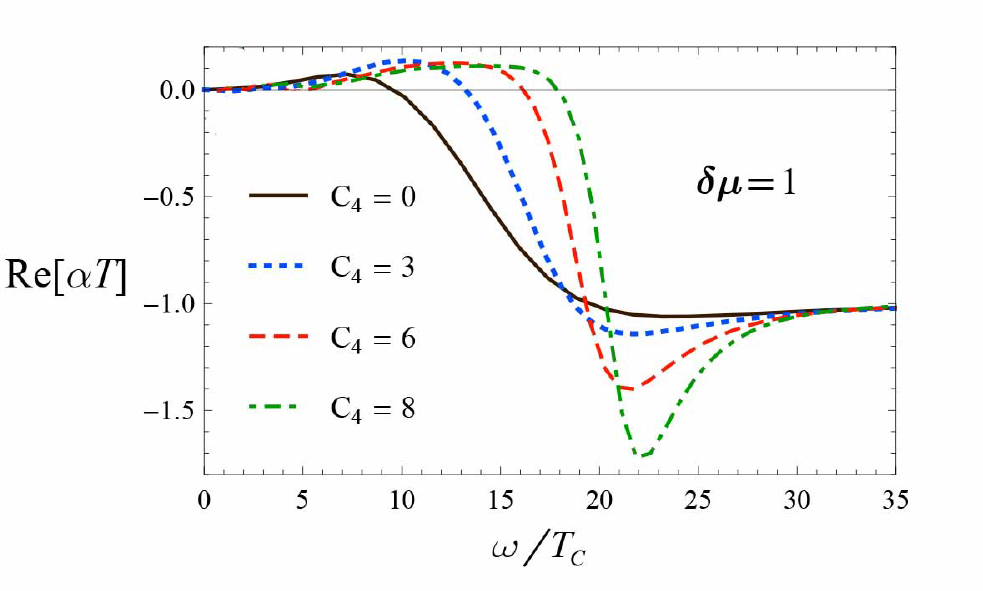}}
      \qquad
        \subfloat[]{\includegraphics[width=0.5\columnwidth]{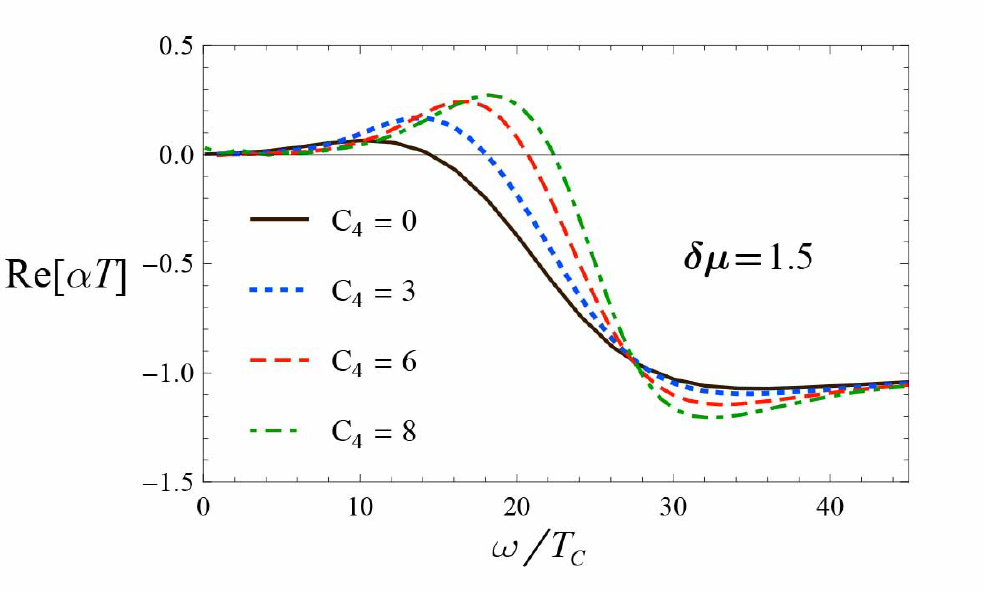}}
           \caption{The real part of the thermo-electric conductivity in terms of $\omega/T_c$ for function ${\cal F}(\psi)=\psi^2+C_4\psi^4$ with $C_4=0,3,6,8$ (solid curve, dotted curve, dashed curve, and dot-dashed curve) and $\delta\mu=0.5,1,1.5$ (figures (a), (b), and (c)).}
 \label{CTaT} 
\end{figure}

\begin{figure}[h]
    \centering
        \subfloat[]{\includegraphics[width=0.5\columnwidth]{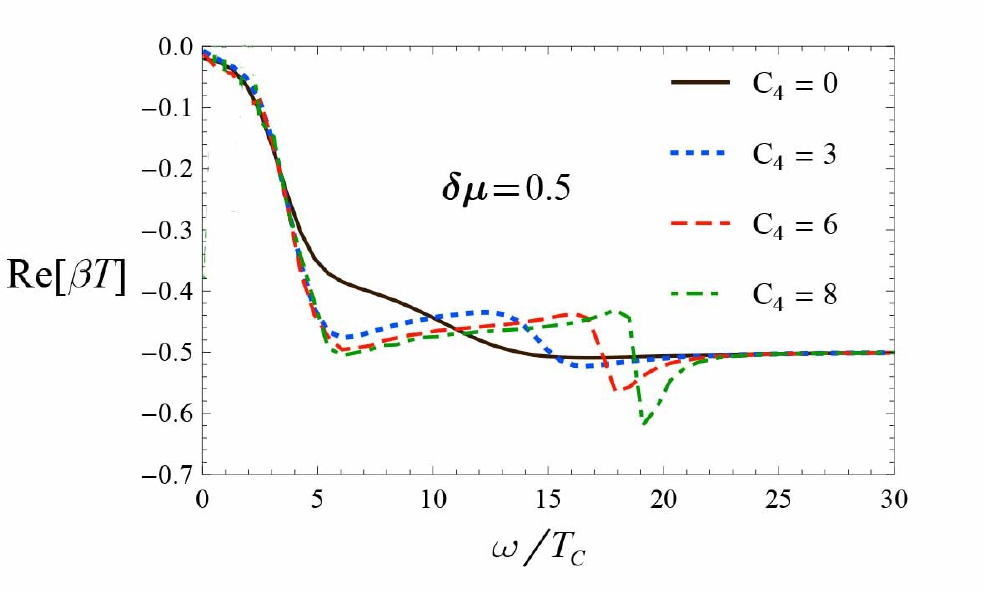}}
        \subfloat[]{\includegraphics[width=0.5\columnwidth]{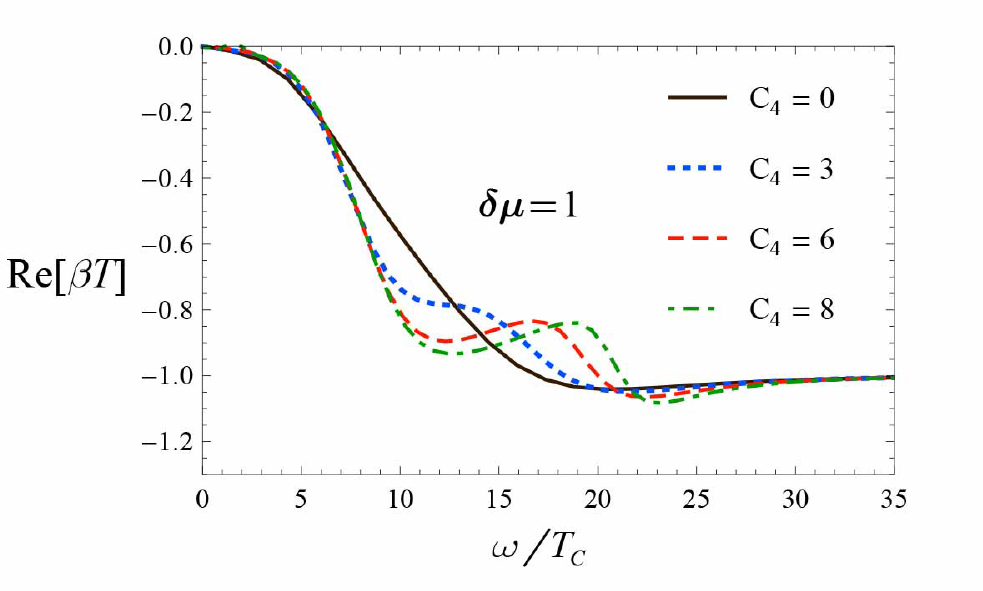}}
     \qquad
        \subfloat[]{\includegraphics[width=0.5\columnwidth]{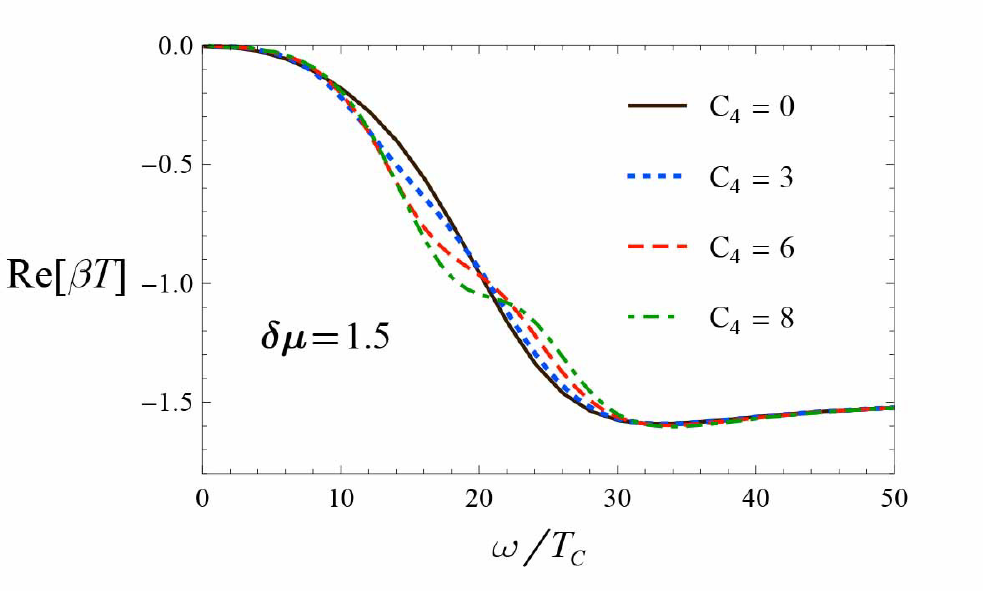}}
         \caption{The real part of the thermo-spin conductivity in terms of $\omega/T_c$ for function ${\cal F}(\psi)=\psi^2+C_4\psi^4$ with $C_4=0,3,6,8$ (solid curve, dotted curve, dashed curve, and dot-dashed curve) and $\delta\mu=0.5,1,1.5$ (figures (a), (b), and (c)).}
     \label{CTbT} 
\end{figure}

\begin{figure}[h]
    \centering
    \subfloat[]{\includegraphics[width=0.5\columnwidth]{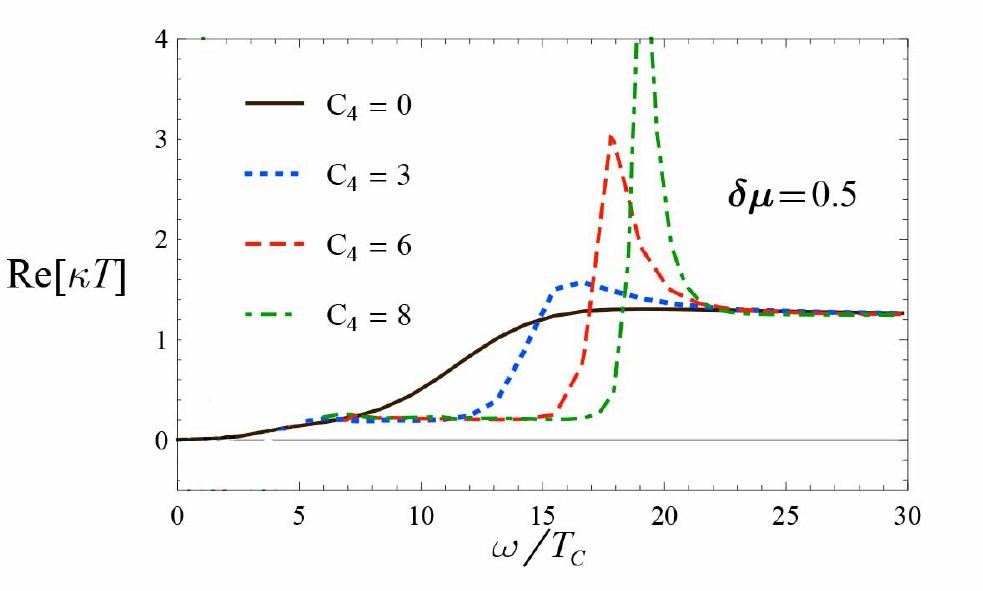}}
    \subfloat[]{\includegraphics[width=0.5\columnwidth]{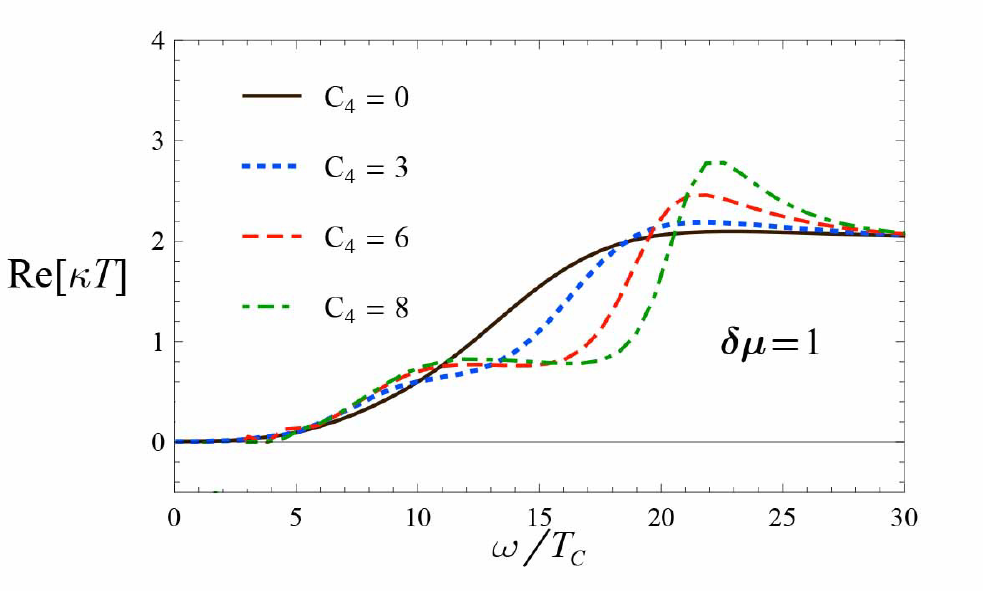}}
     \qquad
        \subfloat[]{\includegraphics[width=0.5\columnwidth]{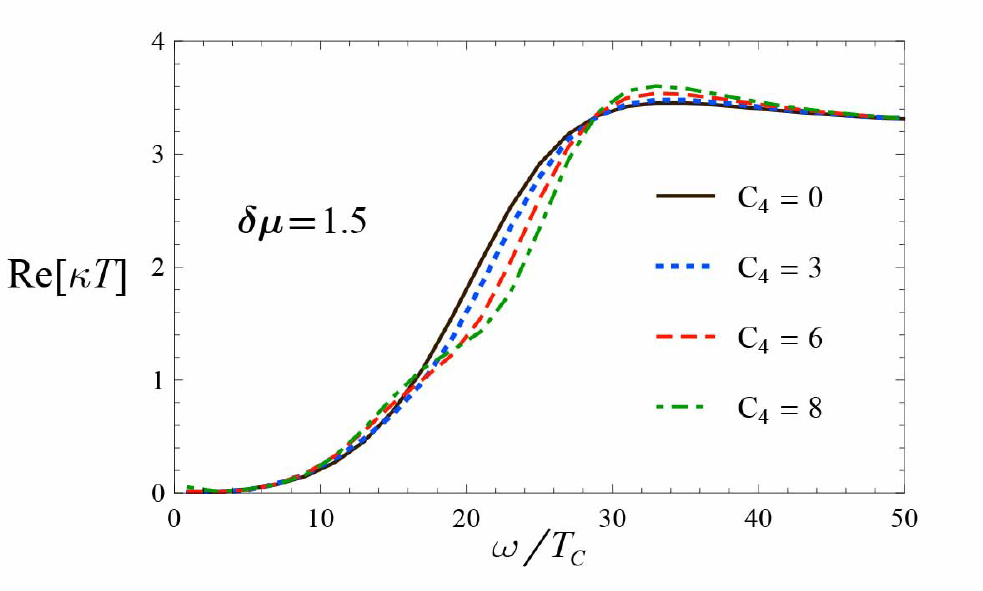}}
           \caption{The real part of the thermal conductivity in terms of $\omega/T_c$ for function ${\cal F}(\psi)=\psi^2+C_4\psi^4$ with $C_4=0,3,6,8$ (solid curve, dotted curve, dashed curve, and dot-dashed curve) and $\delta\mu=0.5,1,1.5$ (figures (a), (b), and (c)).}
            \label{CTkT} 
\end{figure}

The aim of this part of the study is to determine how $C_{4}$ affects the conductivities, for ${\cal F}(\psi)=\psi^2+C_4\psi^4$.
By increasing $C_4$ the pseudo-gap becomes wider and the coherent peak becomes narrower and stronger.
Comparing plots of Fig. (\ref{CTsA}) (and Fig. (\ref{CTsAC4})), one can also observe that parameter $C_4$ gradually loses its control over coherent peak as a system becomes more unbalanced. On the other hand, this parameter keeps making the pseudo-gap wider even in highly unbalanced systems.

As mentioned, increasing coefficient parameter $C_4$ may produce extra delta functions in the pseudo-gap.
This is the direct consequence of the fact that poles of the imaginary part are mapped by the Kramers-Kroning relation to delta functions, just like the case of the delta function at $\omega=0$.
Formation of these extra delta functions leads to create extra resonances. We should expect such resonances since we have the ``vertex'' $\psi^\alpha(\partial p-A)^2$ with $\alpha\ge 3$ providing inelastic scattering \cite{Polchinski:2002jw}.

For the spin conductivity, in contrast to the electric case, $C_4$ play a minor rule in controlling conductivity fluctuations. Fig. (\ref{CTsB}) indicates the optical spin conductivities for two unbalanced systems with $\delta\mu/\mu=1,1.5$. Moreover, increasing both $\delta\mu/\mu$ and $C_4$ results in stronger depletion at low frequencies. 

For the mixed conductivity, the increase of $C_4$ shifts fluctuations to larger frequencies. This happens because of the suppression of negative fluctuations at small $\omega$ and the amplification of positive ones at larger $\omega$, see Fig. (\ref{CTg}). It means that there is a shift in fluctuations towards positive conductivities and this is more noticeable in less unbalanced systems.

In the case of the thermo-electric conductivity, Fig. (\ref{CTaT}) shows that $C_4$ has a remarkable control over the fluctuations. In more unbalanced systems, the growth in $C_4$ intensifies the fluctuations not only in the negative direction but also in the positive direction (at smaller frequencies).
Although the positive fluctuations in more unbalanced systems kills the pseudo-gap, it becomes wider by raising $C_4$ in less unbalanced ones (e.g. systems with $\delta \mu / \mu=0.5$).

Fig. (\ref{CTbT}) shows the appearance of slight thermo-spin conductivity fluctuations caused by increasing $C_4$ at middle frequencies. The fluctuations are also suppressed when a system becomes more unbalanced. 

As one can see from Fig. (\ref{CTkT}), the fluctuations of the thermal conductivity are dominated by increasing $C_4$. Furthermore, the coherent peak gets sharper and shifts towards larger frequencies when $C_4$ grows. But, as evident in Fig. (\ref{CTkT}), it seems that these behaviors vanish in highly unbalanced systems.

\clearpage
%%%%%%%%%%%%%%%%%%%%%%%%%%%%%%%%%%%%%%%%%%%%%%%%%%%%%%%%%%%

\subsubsection{Conductivity behavior with respect to $\alpha$}

\begin{figure}[h]
    \centering
        \subfloat[]{\includegraphics[width=0.5\columnwidth]{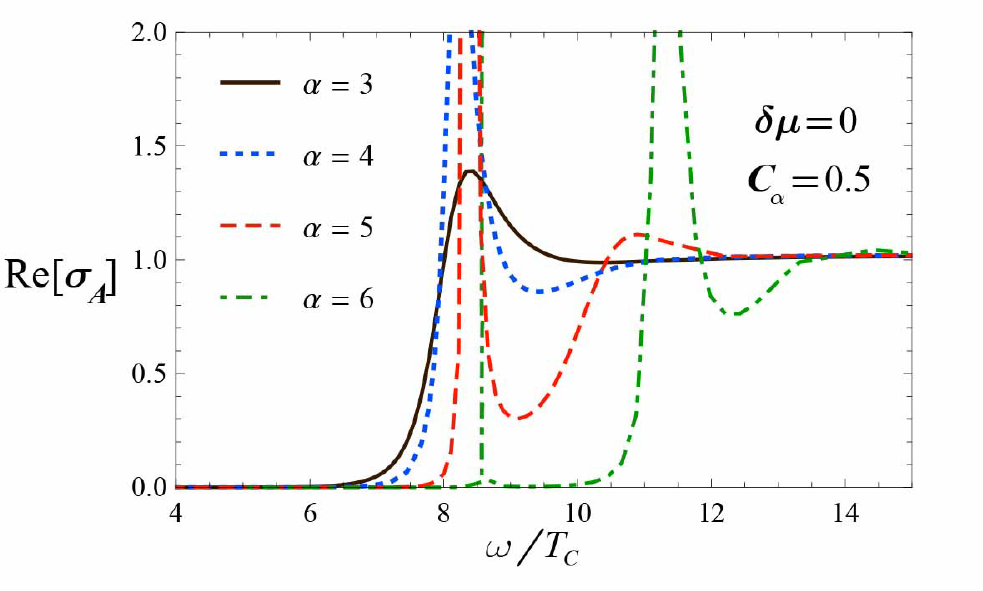}}
        \subfloat[]{\includegraphics[width=0.5\columnwidth]{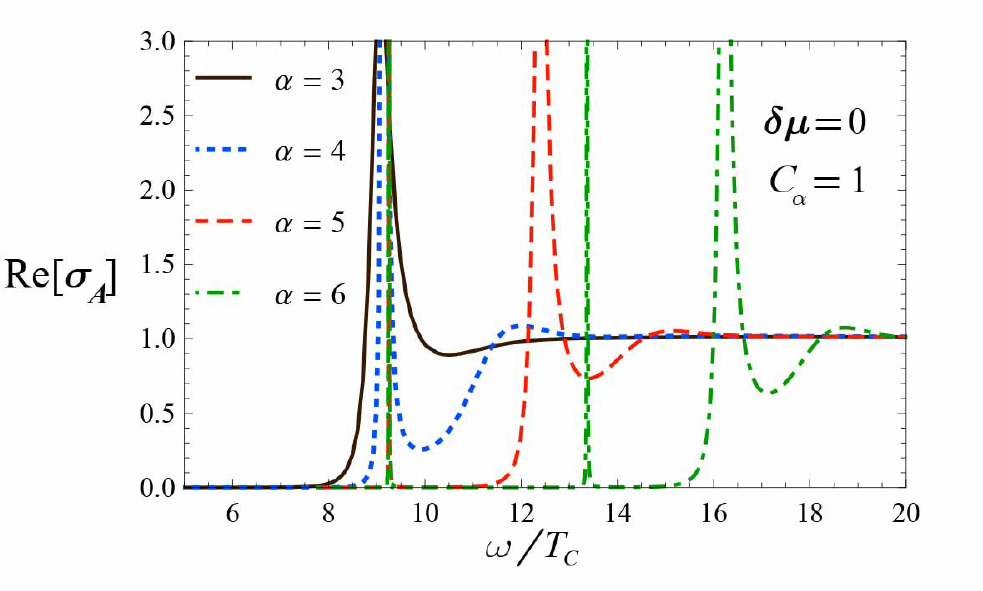}}
      \qquad
        \subfloat[]{\includegraphics[width=0.5\columnwidth]{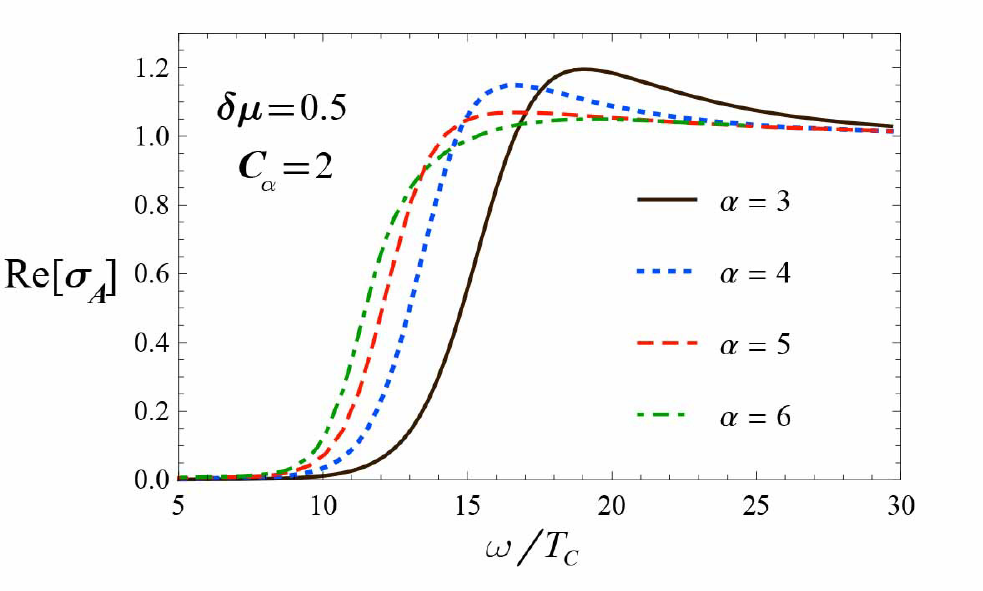}}
        \subfloat[]{\includegraphics[width=0.5\columnwidth]{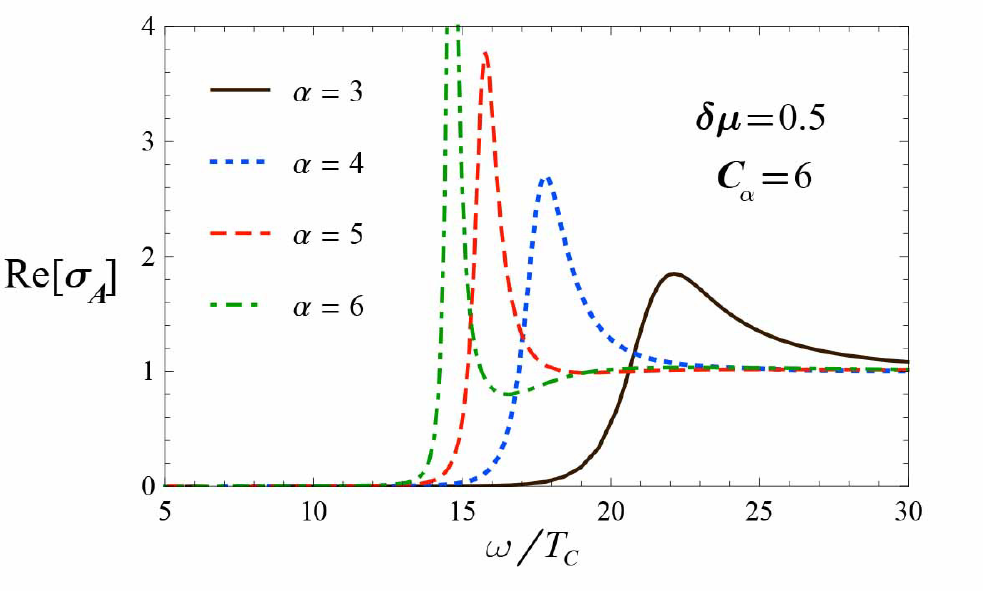}}
      \qquad
        \subfloat[]{\includegraphics[width=0.5\columnwidth]{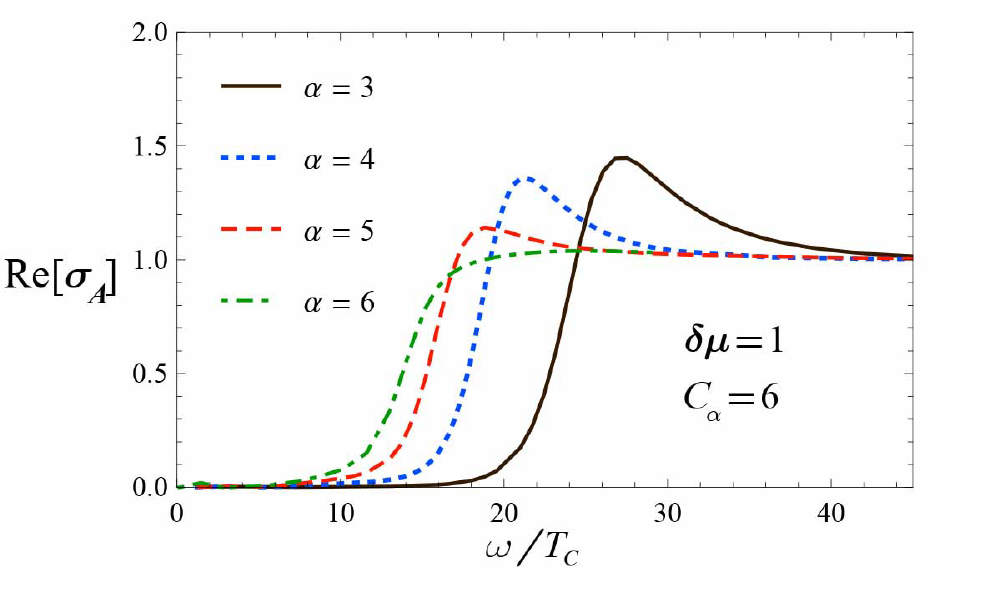}}
        \subfloat[]{\includegraphics[width=0.5\columnwidth]{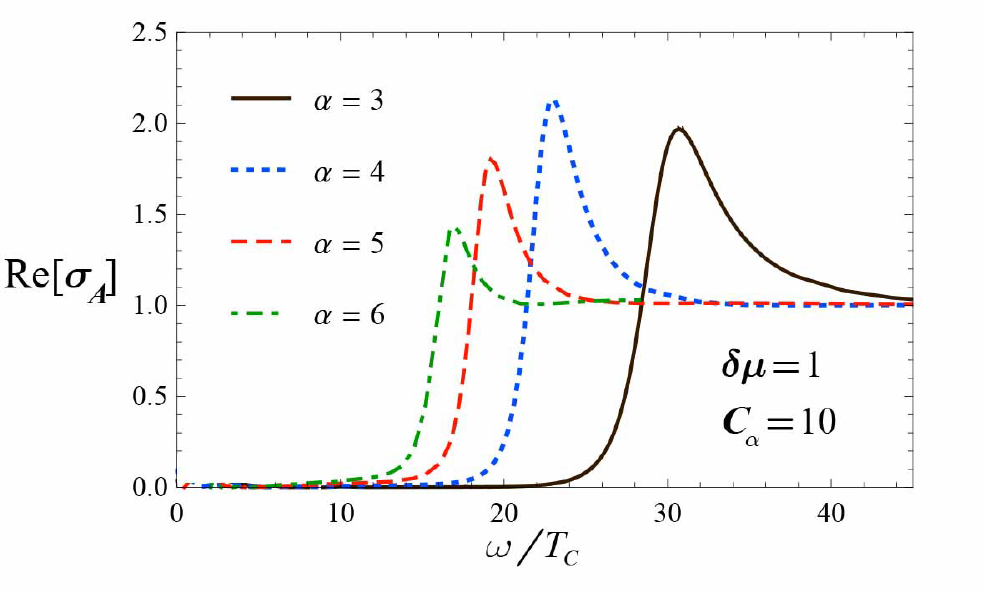}}
      \qquad
        \subfloat[]{\includegraphics[width=0.5\columnwidth]{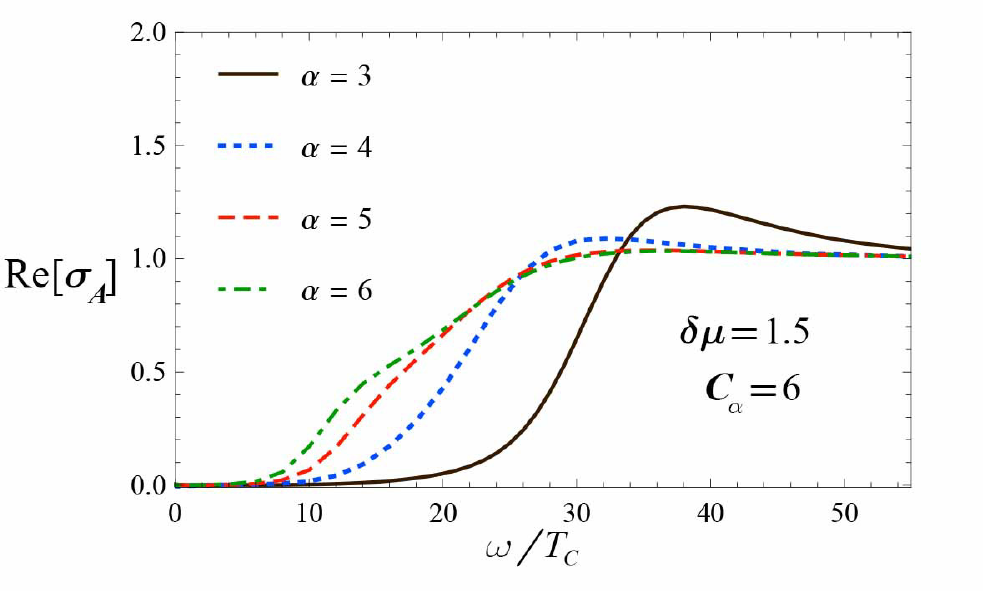}}
        \subfloat[]{\includegraphics[width=0.5\columnwidth]{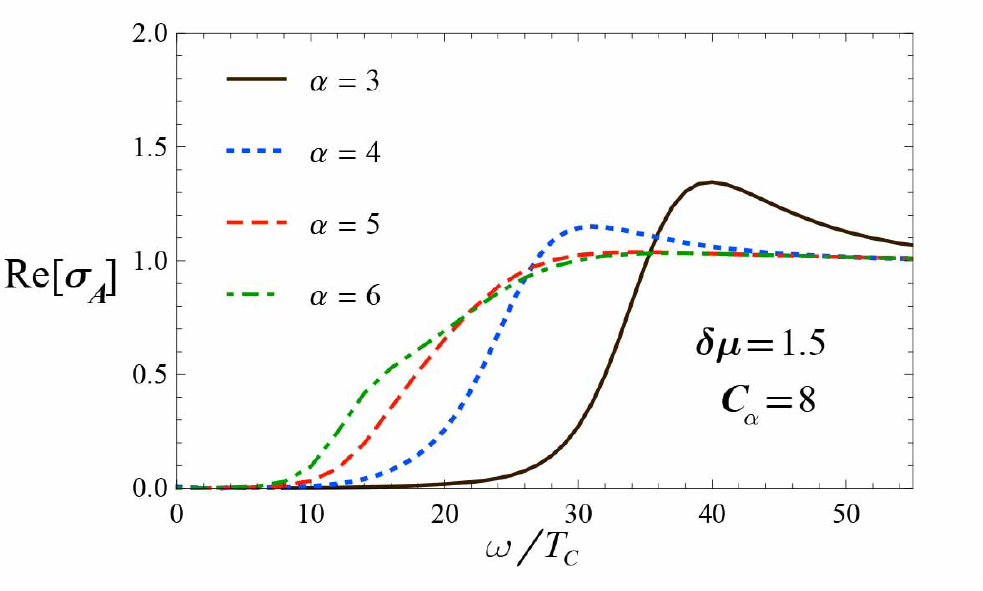}}
          \caption{The optical electric conductivities in terms of $\omega/T_c$ for function ${\cal F}(\psi)=\psi^2+C_\alpha\psi^\alpha$ with $\alpha=3,4,5,6$ (solid curve, dotted curve, dashed curve, and dot-dashed curve). Each row is related to systems with same imbalance; we have $\delta\mu=0,0.5,1,1.5$ from up to down.}
            \label{CTsAa} 
\end{figure}

\begin{figure}[h]
    \centering
        \subfloat[]{\includegraphics[width=0.5\columnwidth]{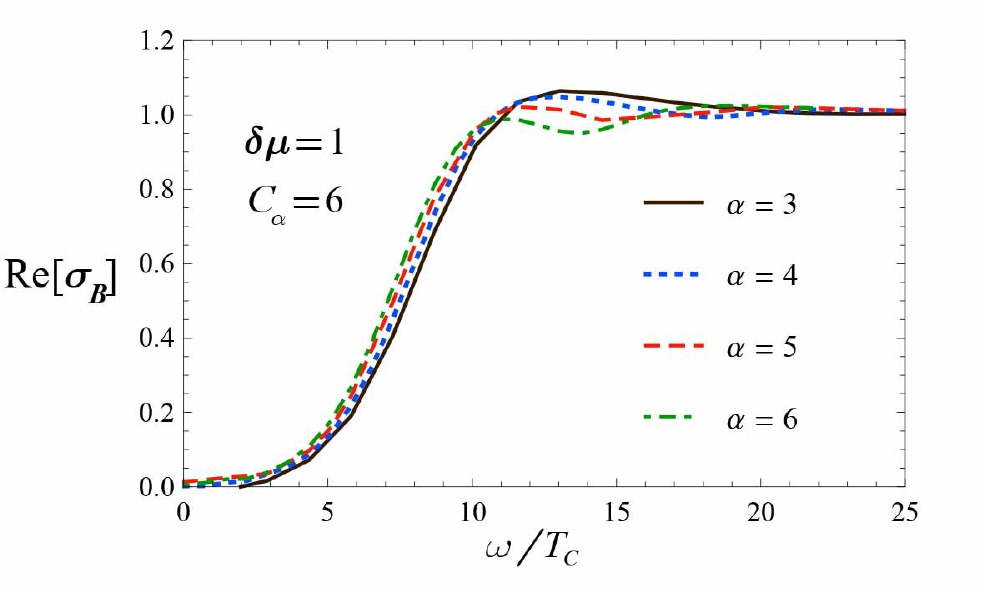}}
        \subfloat[]{\includegraphics[width=0.5\columnwidth]{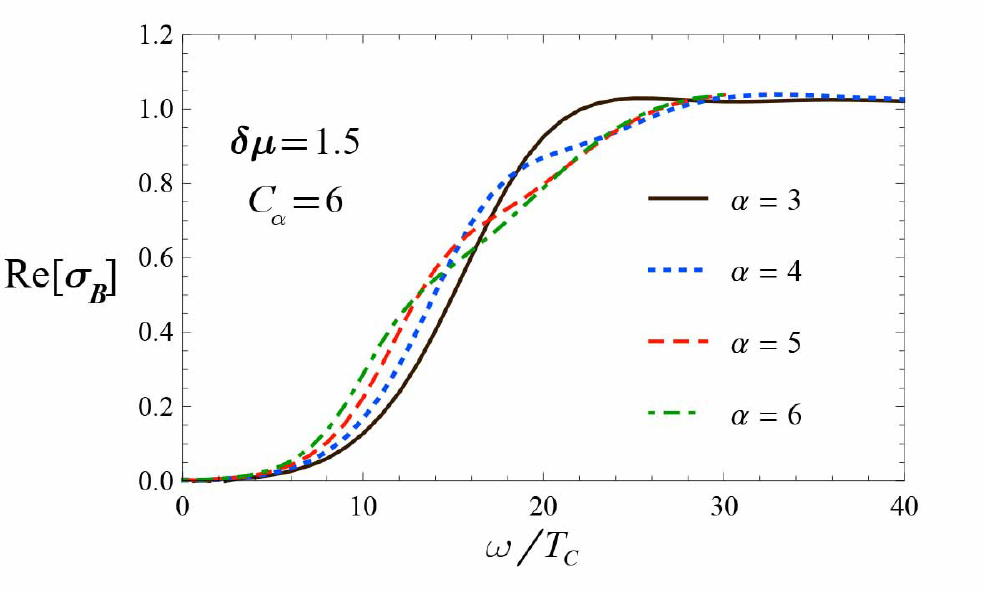}}
           \caption{The optical spin conductivities in terms of $\omega/T_c$ for function ${\cal F}(\psi)=\psi^2+C_\alpha\psi^\alpha$ with $\alpha=3,4,5,6$ (solid curve, dotted curve, dashed curve, and dot-dashed curve) and fixed $C_\alpha=6$. The left and right figures belong to systems with $\delta\mu=1$ and $1.5$.}
      \label{CTsBa}
\end{figure}

\begin{figure}[h]
    \centering
        \subfloat[]{\includegraphics[width=0.5\columnwidth]{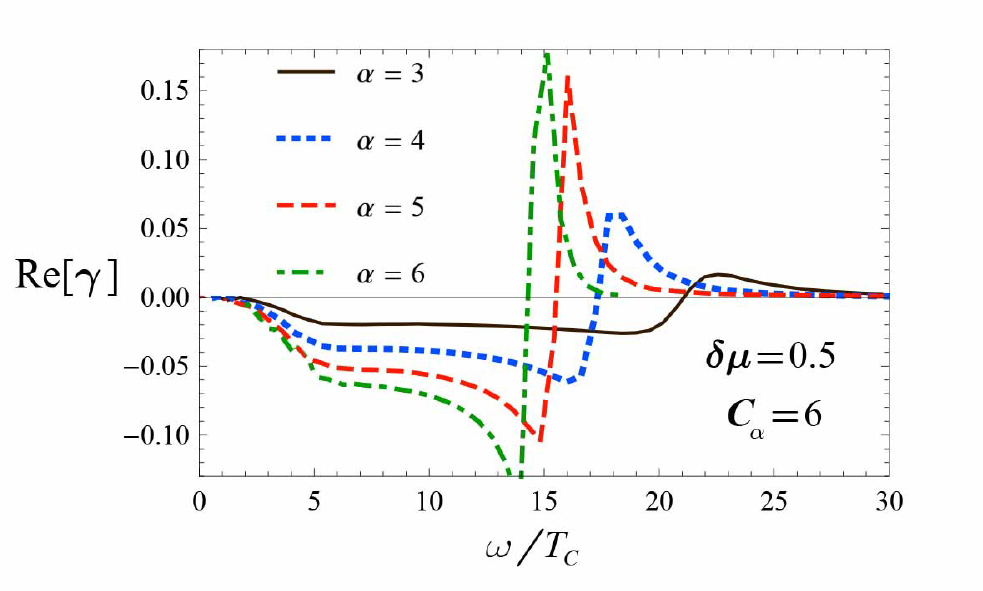}}
        \subfloat[]{\includegraphics[width=0.5\columnwidth]{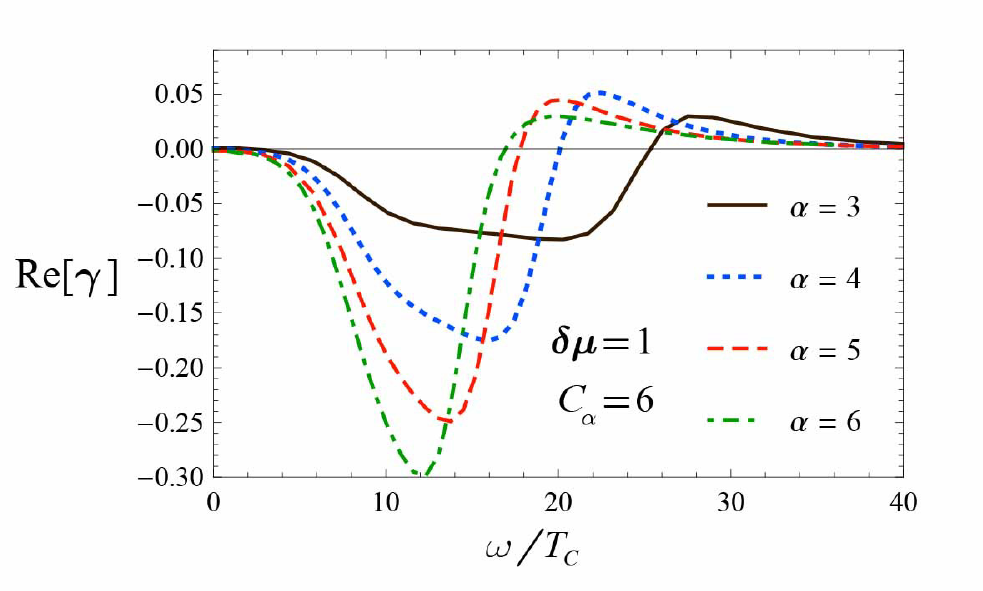}}
        \qquad
        \subfloat[]{\includegraphics[width=0.5\columnwidth]{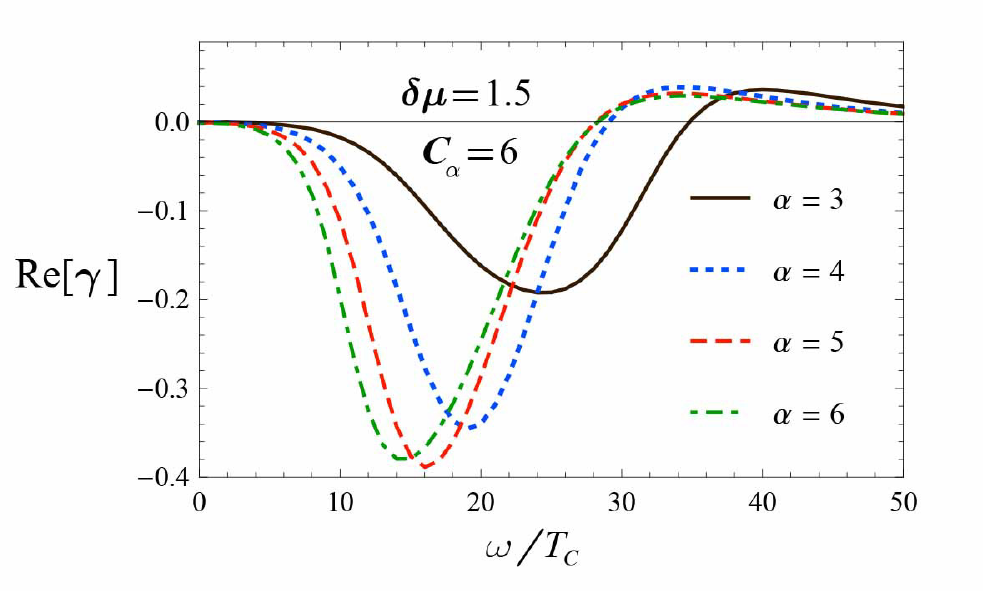}}
          \caption{The real part of mixed conductivities in terms of $\omega/T_c$ for function ${\cal F}(\psi)=\psi^2+C_\alpha\psi^\alpha$ with $\alpha=3,4,5,6$ (solid curve, dotted curve, dashed curve, and dot-dashed curve), $C_\alpha=6$, and $\delta\mu=0.5,1,1.5$ (figures (a), (b), and (c)).}
           \label{CTga}
\end{figure}

\begin{figure}[h]
    \centering
        \subfloat[]{\includegraphics[width=0.5\columnwidth]{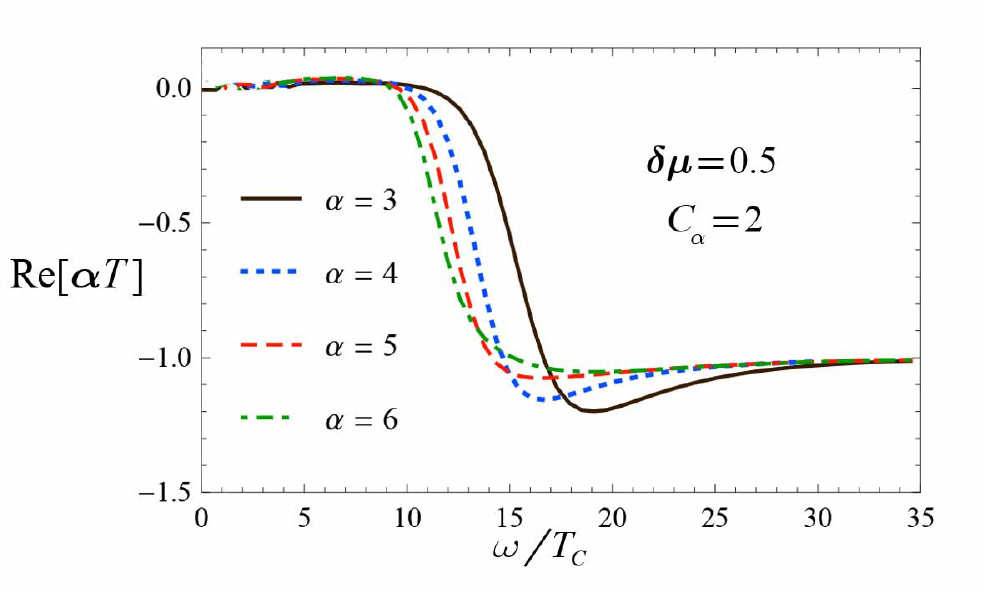}}
        \subfloat[]{\includegraphics[width=0.5\columnwidth]{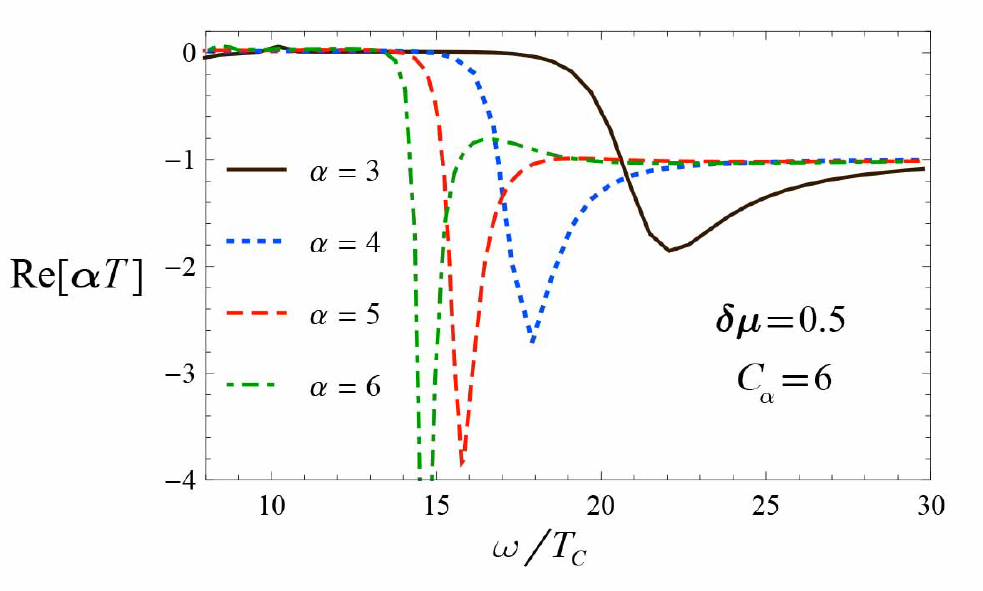}}
      \qquad
        \subfloat[]{\includegraphics[width=0.5\columnwidth]{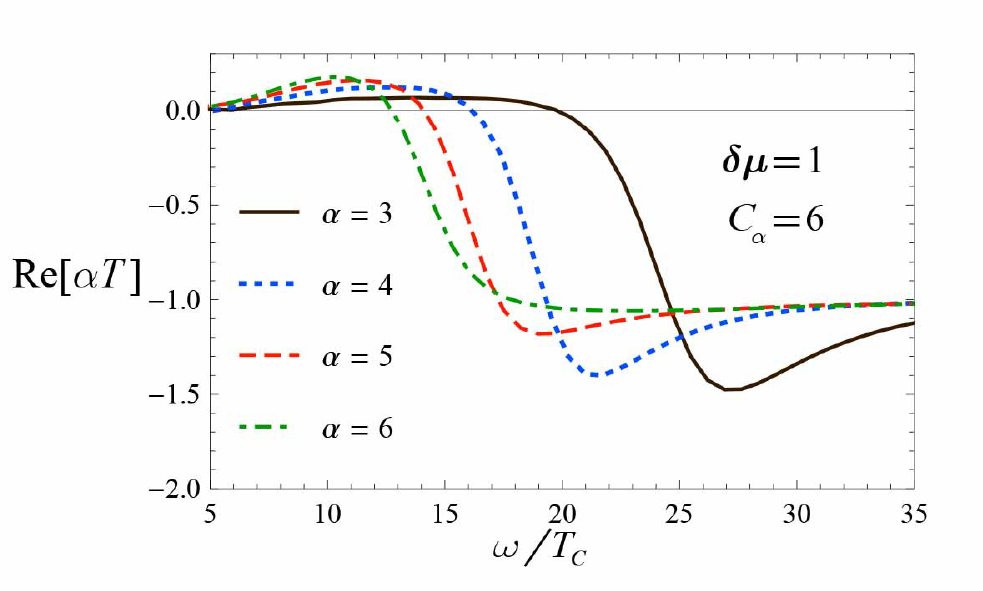}}
        \subfloat[]{\includegraphics[width=0.5\columnwidth]{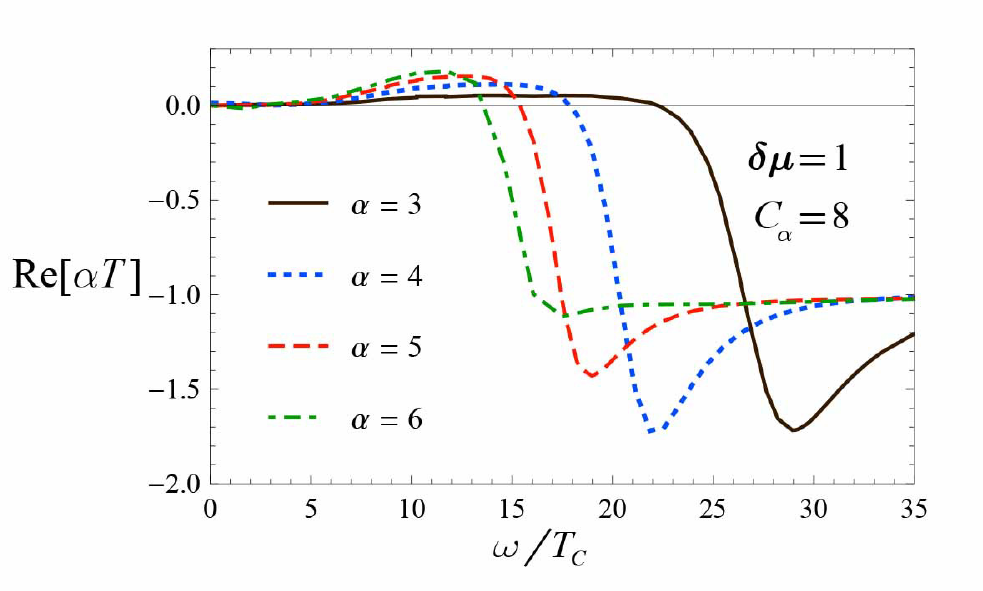}}
      \qquad
        \subfloat[]{\includegraphics[width=0.5\columnwidth]{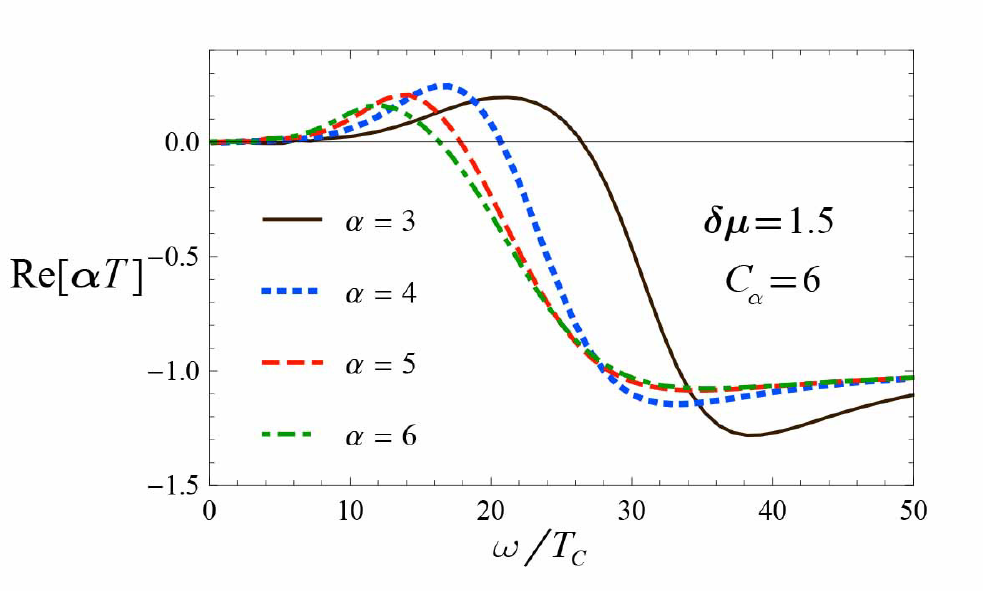}}
        \subfloat[]{\includegraphics[width=0.5\columnwidth]{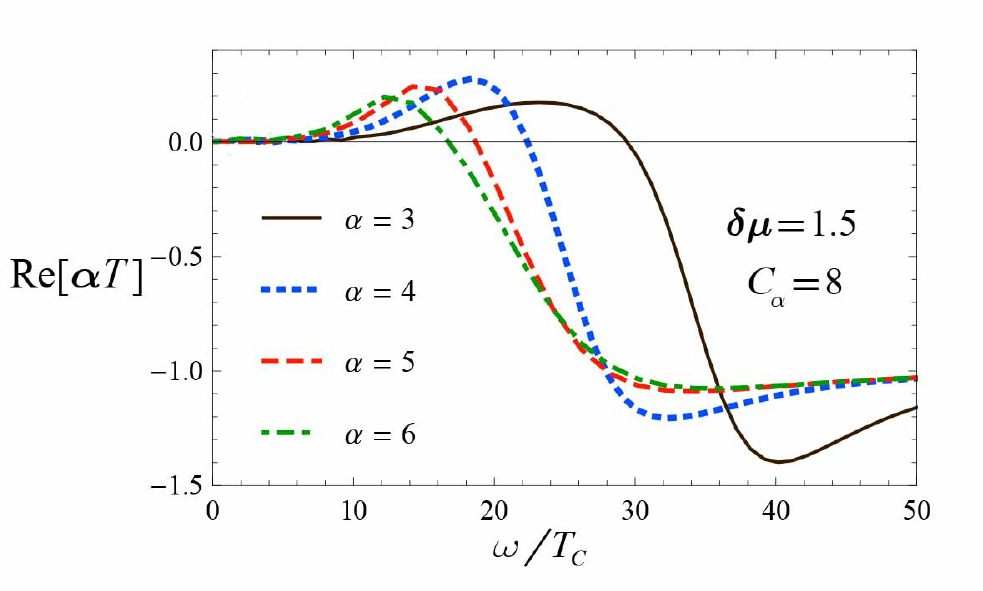}}
           \caption{The real part of thermo-electric conductivities in terms of $\omega/T_c$ for function ${\cal F}(\psi)=\psi^2+C_\alpha\psi^\alpha$ with $\alpha=3,4,5,6$ (solid curve, dotted curve, dashed curve, and dot-dashed curve). Each row is related to the systems with the same imbalance; we have $\delta\mu=0,0.5,1,1.5$ from up to down.}
         \label{CTaTa} 
\end{figure}

\begin{figure}[h]
    \centering
        \subfloat[]{\includegraphics[width=0.5\columnwidth]{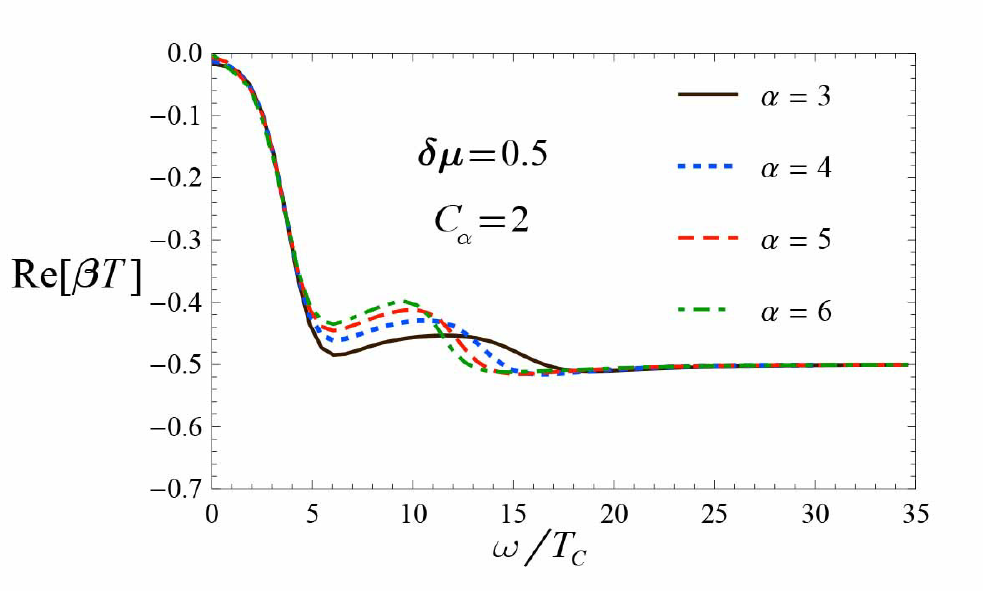}}
        \subfloat[]{\includegraphics[width=0.5\columnwidth]{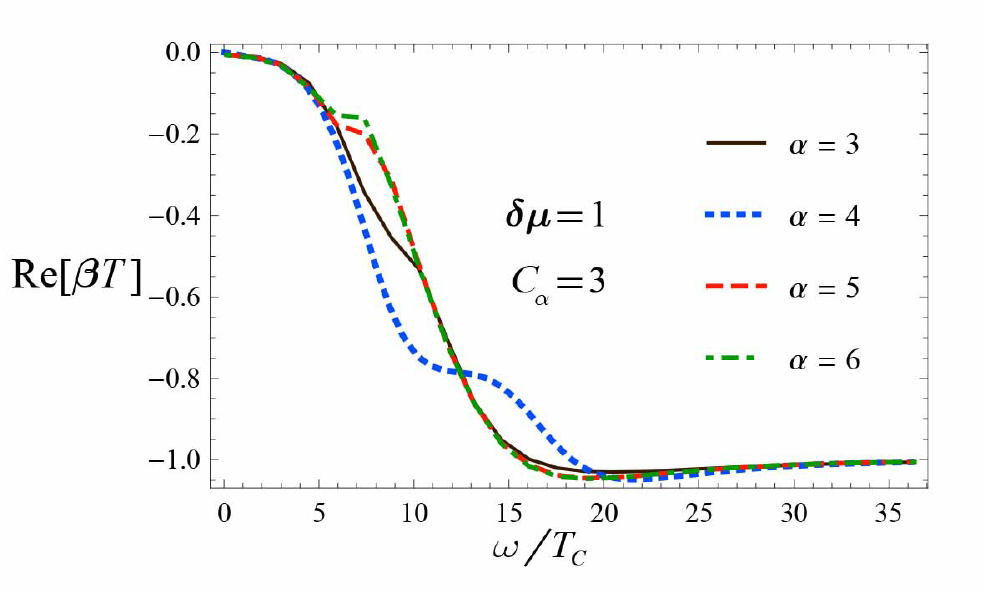}}
      \qquad
        \subfloat[]{\includegraphics[width=0.5\columnwidth]{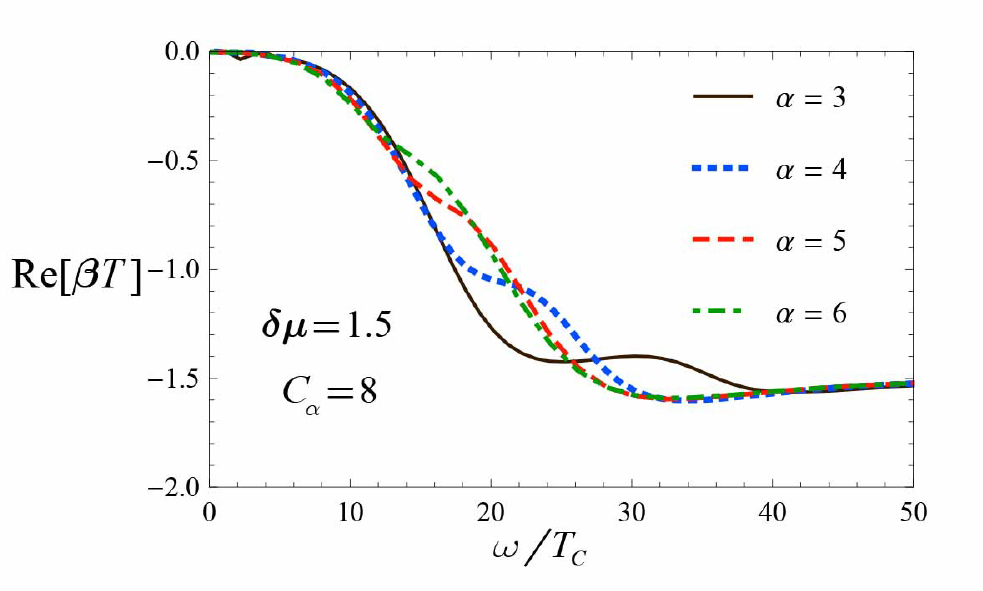}}
          \caption{The real part of the thermo-spin conductivities in terms of $\omega/T_c$ for function ${\cal F}(\psi)=\psi^2+C_\alpha\psi^\alpha$ with $\alpha=3,4,5,6$ (solid curve, dotted curve, dashed curve, and dot-dashed curve).}
            \label{CTbTa}
\end{figure}

\begin{figure}[h]
    \centering
        \subfloat[]{\includegraphics[width=0.5\columnwidth]{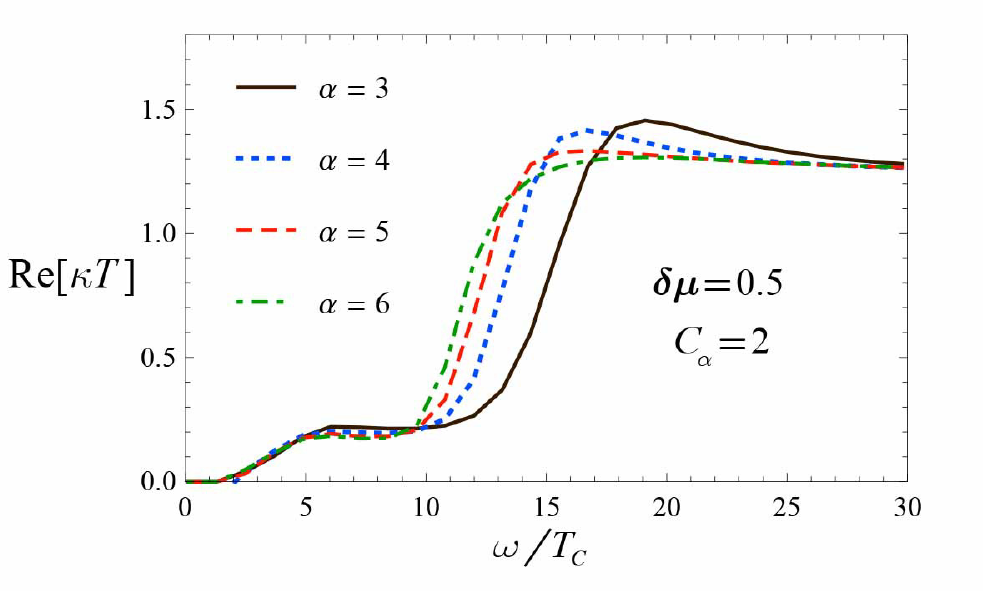}}
        \subfloat[]{\includegraphics[width=0.5\columnwidth]{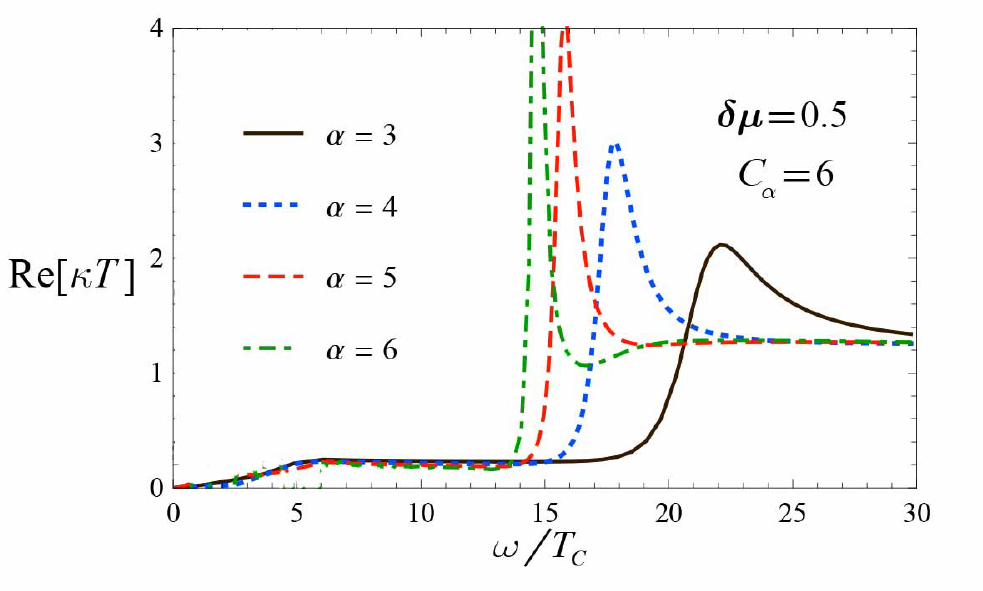}}
      \qquad
        \subfloat[]{\includegraphics[width=0.5\columnwidth]{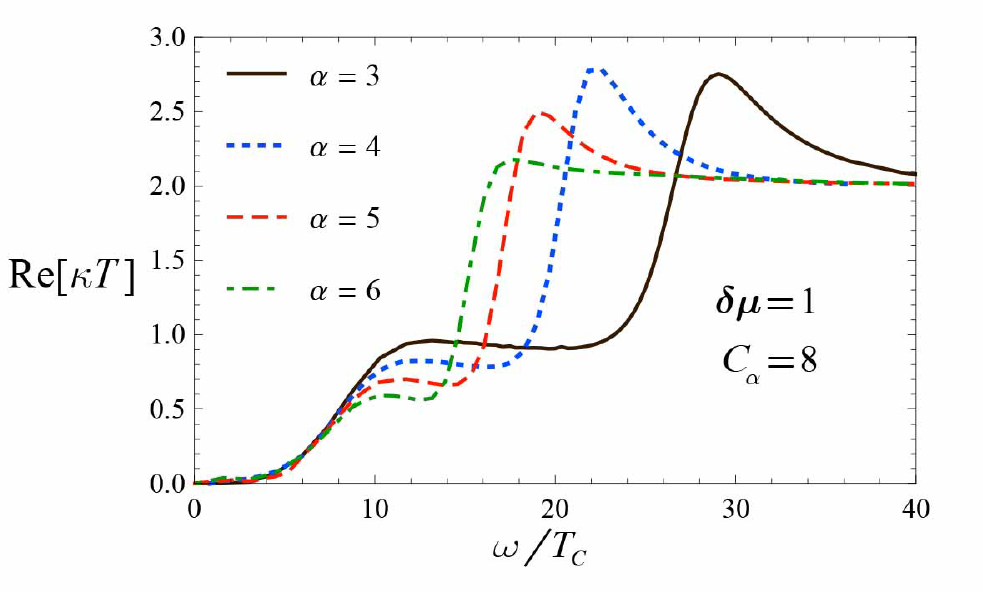}}
        \subfloat[]{\includegraphics[width=0.5\columnwidth]{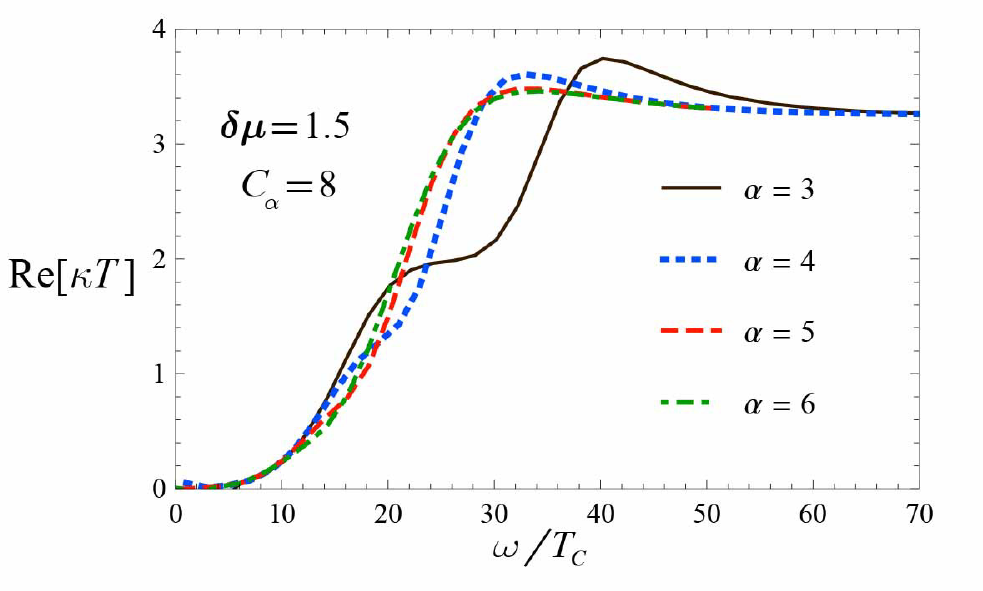}}
          \caption{The real part of the thermal conductivity in terms of $\omega/T_c$ for the function ${\cal F}(\psi)=\psi^2+C_\alpha\psi^\alpha$ with $\alpha=3,4,5,6$ (solid curve, dotted curve, dashed curve, and dot-dashed curve).}
      \label{CTkTa} 
\end{figure}

By assuming the parameter $C_\alpha$ to be fixed, we can study how conductivities behave when $\alpha$ varies.
From Fig. (\ref{CTsAa}) part (c) to (h), one can find that conductivity pseudo-gap of an unbalanced system becomes smaller by the growth of $\alpha$.

In balanced systems, the part (a) of Fig. (\ref{CTsAa}) obviously shows that, in balanced systems, both $\alpha$ and $C_\alpha$ control the strength of the fluctuations \cite{Franco:2009if}. Although growth of the $\alpha$ parameter makes the coherent peak of the optical electric conductivity sharper and higher in balanced systems, it is not what always happens in the case of unbalanced ones.
For instance, in Figs. (\ref{CTsAa}) (c) and (e)-(h), one can see the suppression of fluctuations when  $\alpha$ increases. It is interesting that for $\delta\mu/\mu=0.5$ quite opposite treatments in the fluctuations can be observed when $C_\alpha=2$ and $C_\alpha=6$.
Therefore, how the fluctuations are affected by increasing $\alpha$ depends on the value of $C_\alpha$.
Observe that, for $\delta\mu/\mu=1$ and $1.5$, Figs. (\ref{CTsAa}) (f) and (h) illustrate damped fluctuations even for large $C_\alpha$s. Indeed, we could not find a large enough $C_\alpha$ for which $\alpha$ amplifies the fluctuations (it does not happen even for $C_\alpha=14$).
As an interesting result, changes in $C_\alpha$ and $\alpha$ values can approach $\omega_g/T_c$ to $8$ which is similar to the standard holographic superconductor model \cite{Hartnoll:2008vx}. For example, in $\delta\mu/\mu=1.5$ and $C_\alpha=8$, the $\omega_g/T_c$ ratio goes to about $8$ by setting $\alpha=6$ (Fig (\ref{CTsAa}) (h)).

Fig. (\ref{CTsBa}) displays that for large $\alpha$, there is a slight reduction of the optical spin conductivity pseudo-gap and reinforcement of the fluctuations. It means that spin conductivity is not as sensitive as the other conductivity types to the parameter $\alpha$.

However, in Fig. (\ref{CTga}), the mixed conductivity plots show an increase in strength of the fluctuations as long as $\alpha$ grows. 
The influence of $\alpha$ on the mixed conductivity of balanced systems is stronger compared with unbalanced ones.
As shown in Fig. (\ref{CTaTa}), there is a movement in the fluctuations towards smaller frequencies for large $\alpha$ values in the thermo-electric conductivity. Similar to the electric case, whether the increase of $\alpha$ intensifies fluctuations or not depends on the values of both $C_\alpha$ and the imbalance. Fig. (\ref{CTaTa}) (b) shows that, for the case of $\delta \mu/\mu=0.5$, the parameter $C_\alpha=6$ is large enough to has the fluctuations amplified by increasing $\alpha$. Nevertheless, for highly unbalanced systems with $\delta \mu=1$ and $1.5$, even for $C_\alpha=8$, the fluctuations are suppressed by increasing $\alpha$ (Fig. (\ref{CTaTa}) (a) and (c)-(f)).

Moreover, in the case of the optical thermo-spin conductivity, one can figure out from Fig. (\ref{CTbTa}) that the growth of the $\alpha$ parameter produces slight fluctuations. Similar to the spin conductivity, these fluctuations do not obey an explicit pattern, but they are damped by increasing the imbalance.
Observe that we need larger $C_\alpha$ to well demonstrate the conductivity fluctuations of more unbalanced systems.

The thermal and electric conductivities behave with varying the $\alpha$ parameter in almost the same manner. For the balanced case, the real part of the thermal conductivity reduces to the optical electric one.
The growth of the $\alpha$ parameter is generally followed by a shift of the conductivity fluctuations and the coherent peak towards lower frequencies, while their amplification depends on the values of both the imbalance and $C_\alpha$.
According to Fig. (\ref{CTkTa}) (a) and (b), there exist two opposite behaviors for two different values of $C_\alpha$ in the system with $\delta\mu/\mu=0.5$. Nevertheless, in more unbalanced systems, fluctuations are damped in our range of parameter $C_\alpha$ (even for $C_\alpha=10$ in the system of $\delta\mu/\mu=1$).

\clearpage
%%%%%%%%%%%%%%%%%%%%%%%%%%%%%%%%%%%%%%%%%%%%%%%%%%%%%%%%%%%%%%%%%%%
\section{Conclusion}\label{sec4}
We studied the unbalanced holographic superconductor model in combination with St\"{u}ckelberg mechanism which gives us a highly flexible dual theory. This flexibility gives us more freedom in tuning the model parameters with experiments.
We showed that while model parameter $C_4$ provides the change of phase transition order from second to first, the imbalance makes it harder. In other words, we need a larger $C_4$ for a more unbalanced system to change the order of phase transition. 
Such behavior also can be observed from conductivity diagrams.
In most cases, conductivities behavior of highly unbalanced systems is less influenced by the St\"{u}ckelberg model parameters compared with the less unbalanced ones. In other words, St\"{u}ckelberg mechanism generally loses its effects as system becomes more unbalanced. 
Moreover, we have numerically recovered the Eq. \eqref{beta2} also for the case of unbalanced system.

Additionally, we have found that imbalance can significantly divert the system's behavior with model parameters of St\"{u}ckelberg mechanism. Such deviations can even reverse the behavior in some cases. We can specifically mention the behavior of electric and thermal conductivity with model parameter $\alpha$. For example, electric conductivity fluctuations of the relatively less unbalanced system with $\delta\mu/\mu=0.5$ are intensified as $C_4=6$ although they are damped as $C_4=2$. The same has been also observed in the case of thermal conductivity.

It is interesting to investigate inhomogeneous superconductors in our model. However, we do not observe a Chandrasekhar-Clogston-like bound \cite{Chandrasekhar,Clogston} at zero temperature (for our choice of the model parameters and function (\ref{F})). Therefore, the LOFF phase is not expected to occur in our model. Nevertheless, different choices of model parameters may allow for Chandrasekhar-Clogston-like bounds.

As a future task, we should push more towards the experimental directions and comparisons by making use of the method introduced in \cite{Amoretti:2015gna}. It would be interesting to take advantage of the freedom of ${\cal F}$ to simultaneously match two phenomenological behavior, i.e. the phase transition and conductivity.

\section*{Acknowledgments}

We would like to thank Daniele Musso for helpful discussions on the numerical methods and F. Lalehgani Dezaki for comments on the manuscript.
AJH wishes to thank the Shahrood University of Technalogy and IPM for hospitality during the course of this work.
We would like to thank the referee for his/her instructive comments.
%%%%%%%%%%%%%%%%%%%%%%%%%%%%%%%%%%%%%%%%%%%%%%%%%%%%%%%%%%%%%%%%%%%%%

\end{document}